\documentclass[AMA,STIX1COL]{WileyNJD-v2}

\usepackage{moreverb}
\usepackage{amsmath}
\usepackage{caption} 
\usepackage{subcaption}
\usepackage{graphicx}
\usepackage{pgfplots}
\usepackage[all]{nowidow}
\usepackage[utf8]{inputenc}
\usepackage{multicol}
\usepackage{hyperref}

\newcommand\BibTeX{{\rmfamily B\kern-.05em \textsc{i\kern-.025em b}\kern-.08em
T\kern-.1667em\lower.7ex\hbox{E}\kern-.125emX}}

\articletype{Article Type}

\received{<day> <Month>, <year>}
\revised{<day> <Month>, <year>}
\accepted{<day> <Month>, <year>}

\begin{document}

\title{Joint Quantile Disease Mapping with Application to Malaria and G6PD Deficiency}

\author[1,2]{Hanan Alahmadi*}

\author[1]{H{\aa}vard Rue}

\author[1]{Janet van Niekerk}

\authormark{Alahmadi, H \textsc{et al}}

\address[1]{CEMSE Division, King Abdullah University of Science and Technology, Kingdom of Saudi Arabia}
\address[2]{King Saud University, Kingdom of Saudi Arabia}

\corres{*Hanan Alahmadi. \email{hanan.alahmadi@kaust.edu.sa}}

\abstract[Abstract]{Statistical analysis based on quantile regression methods is more comprehensive, flexible, and less sensitive to outliers when compared to mean regression methods. When the link between different diseases are of interest, joint disease mapping is useful for measuring directional correlation between them. Most studies study this link through multiple correlated mean regressions. In this paper we propose a joint quantile regression framework for multiple diseases where different quantile levels can be considered. We are motivated by the theorized link between the presence of Malaria and the gene deficiency G6PD, where medical scientist have anecdotally discovered a possible link between high levels of G6PD and lower than expected levels of Malaria initially pointing towards the occurrence of G6PD inhibiting the occurrence of Malaria. This link cannot be investigated with mean regressions and thus the need for flexible joint quantile regression in a disease mapping framework. Our joint quantile disease mapping model can be used for linear and non-linear effects of covariates by stochastic splines, since we define it as a latent Gaussian model. We perform Bayesian inference of this model using the INLA framework embedded in the R software package \textit{INLA}. Finally, we illustrate the applicability of model by analyzing the malaria and G6PD deficiency incidences in $21$ African countries using linked quantiles of different levels.}


\maketitle

\section{Introduction}
Malaria is considered a leading cause of mortality worldwide, and the disease is most prominent in Africa. It has been estimated that malaria in 2017 affected about 219 million people and causing around 435,000 deaths \cite{talapko2019malaria}. The Malaria Atlas Project \cite{atlasmalaria} provides a global database on malaria risk in order to solve critical questions. This project disseminates free, accurate, and up-to-date geographical on malaria and associated topics. One of their research outputs points out a relationship between malaria and Glucose 6 phosphate dehydrogenase (G6PD) deficiency, a genetic disorder that affects red blood cells. The G6PD is a gene that provides instructions for making the glucose-6-phosphate dehydrogenase enzyme. The research by the Malaria Atlas Project found that G6PD deficiency is common in populations that have a high level of malaria infection \cite{howes2012G6PD}. Studies dating back to the early 1960s, \cite{allison1960,allison1961malaria}, postulated that G6PD deficiency inhibits the occurrence of malaria. The reasoning was that G6PD deficiency leads to the accumulation of oxygen radicals inside the red blood cells ($H_{2}O_{2}$). This accumulation offers resistance against malaria infection because the \textit{Plasmodium falciparum} parasite (the parasite that causes malaria) does not have any antioxidant mechanism, which makes them more vulnerable to oxygen radicals \cite{aziz2019biochemistry,laslett2021glucose}. The hypothesis that G6PD deficiency provides some protection against \textit{Plasmodium falciparum} malaria was further supported by a review by Greene \cite{greene1993}, published in 1993, based on experimental and population studies. At the same time, it was acknowledged that there is not enough data in population studies, due to limited sample sizes, to produce concluding evidence \cite{beutler1994G6PD,greene1993}. However, there are opposing arguments, also based on limited population studies, stating that G6PD deficiency by itself is unlikely to produce a significant protection against malaria, see \cite{martin1979severe}. In 1995, Ruwende et al. \cite{Ruwende1995} suggest, from two case-control studies of more than 2,000 African children, that G6PD deficiency reduced the risk of severe malaria by around $50\%$. In 2017, a systematic review by Mbanefo et al. \cite{revg6pdmalar2017} based on a selection of 28 various studies arrived at that G6PD deficiency could potentially offer some protection against uncomplicated malaria, but less likely so for severe malaria.\\

Following the results of \cite{allison1961malaria,beutler1994G6PD}, it is of interest to perform a statistical inference of such a relationship between diseases and quantify the uncertainties involved. For this case, we propose using a quantile-based joint model, instead of the standard joint that models correlation of the means of the two diseases, since the G6PD deficiency may act as a resistance factor against malaria, while not the other way around. Thus, to identify possible directional correlation, this study looks at the joint quantiles between the two diseases by modeling the high quantile of G6PD deficiency and the low quantile of malaria. The joint quantile model can be applied to other disease mapping problems. Quantile regression was introduced by Koenker and Basset \cite{art0}. After that the quantile regression has been widely used, in particular for Bayesian spatial analysis \cite{art7}. The R package bayesQR proposed by \cite{art1} can be used to estimate the parameters in quantile regression using a Bayesian approach with the asymmetric Laplace distribution. This package supports both continuous-dependent and binary-dependent variables. In \cite{art2}, the authors proposed using a negative-binomial regression $\alpha$-quantiles approach with an ecological regression model with application to disease mapping of lip cancer.\\ \\
The main difference to our work is that we consider joint quantile regression with two diseases, instead of a single one. In joint quantile regression, one can model spatial dependence through a Gaussian or t-copula process of the quantile levels \cite{art3}, which could provide certain benefits for cases with heavy-tailed spatial data. One of the approaches to spatial quantile regression is to use the Asymmetric Laplace Process (ALP) for modeling the data \cite{art4}. However, this assumes the data is coming from the ALP, regardless of the actual generating distribution of the data. A quantile regression-based Bayesian joint modeling analysis of longitudinal-survival data has been proposed in \cite{art11} and it extends the use of the asymmetric Laplace process as in \cite{art4} to joint quantile regression. Markov chain Monte Carlo (MCMC) methods have been used for parameter estimation in Bayesian quantile regression models, for example in \cite{art8} for multivariate quantile regression. However, we advocate the use of INLA over MCMC for practical disease mapping due to its computational advantages. Spatial quantile regression is widely used with applications ranging from modeling of wildfire risk \cite{art3} to studying healthy life years expectancy \cite{art9} to economics \cite{art4}. In \cite{art10}, a Bayesian multiple quantile regression method is proposed for linear models, and they used the working likelihood instead of the likelihood of the generating distribution. In contrast, the quantile regression in \cite{padellini2018model} was developed such that the likelihood of the generating distribution is respected. For a comprehensive introduction to quantile regression for spatial data, see \cite{art6}, and for multivariate disease mapping modeling we refer to \cite{art5}, which includes many practical exercises and examples, often provided with R-code implementation. As far as the authors are aware, there is no available literature on joint quantile disease mapping is available, which we aim to contribute in this study.

\section{Disease Mapping}\label{sec:dis_map}
Disease mapping, also known as spatial epidemiology, analyzes the incidence of disease using geographical information. In other words, Disease mapping describe the spatial variation of disease. The two characteristics of disease mapping are the location of the events, which is called spatial or geographical distribution, and the disease.\cite{lawson2018bayesian}. The Poisson distribution is well representing the disease count for the data that have low disease count for a relatively large population \cite{lawson2018bayesian}. For the
region that consists of $n$ non-overlapping areas \cite{riebler2016intuitive}, let $y_{i}$ denote the number of cases in regions $i$. Often $y_{i}$ is assumed to be distributed as :
\begin{align}\label{eq:1}
y_{i} \sim \text{Poisson} (\mu_{i})
\end{align}
where $\mu_{i}$ is the mean of $y_{i}$. The mean function often consists of two components. The first component is usually called the relative risk, which represents the risk within a region, it is unknown and the purpose of this work to estimate these values. The second component is usually called standardization, which represents the expected local count. The expected local count is the value that represents our expectation if the population locally behaved the way the standard population behaves. The expectation of the cases in region $i$ can be written as follows:
\begin{align}\label{eq:2}
E(y_{i}) = \mu_{i} = E_{i} \lambda_{i}
\end{align}
where $E_{i}$ is the expected number for the $i$th area, which is usually assumed to be a fixed quantity \cite{lawson2018bayesian}. The expected number can be obtained by using indirect standardization as follows:
\begin{align}\label{eq:3}
E_{i} = \sum_{j = 1 }^{m} r_{j}^{(s)} n_{j}^{(i)}
\end{align}
here, $r_{j}^{(s)}$ denotes the disease rate of the standard population in stratum j, the rate is the number of cases divided by the population, $n_{j}^{(i)}$ is the size in stratum j of area $i$ \cite{moraga2018small},
and $\lambda_{i}$ is the relative risk for $i$th area. Here  $\lambda_{i} = 1$ means there is no augmented risk in comparison with the whole study area; $\lambda_{i} >  1$ , $\lambda_{i} <  1$ indicates higher risk and lower risk than the average  respectively \cite{blangiardo2013spatial}. The maximum likelihood estimator of $\lambda_{i}$ is $\hat{\lambda_{i}} = y_{i} /E_{i}$ which is correspond to the standardised mortality ratio (SMR). However, mapping SMRs directly are misleading and insufficient for counties with small populations. Therefore the covariates need to be incorporated in order to smooth extreme values because of the small sample sizes by borrowing information from neighboring counties. The model considered in this work for disease mapping is formulated as follow:
\begin{align} \label{eq:4}
y_{i} \sim \text{Poisson}(E_{i} \lambda_{i}) , i = 1,...,n
\end{align}
\begin{align} \label{eq:5}
\log(\lambda_{i}) = \log(\eta_{i}) =  m_{0}+ \sum_{f = 1}^{F} \beta_f X_{i f} + \sum_{r=1}^{R} \rho^r(u_{i r}) + b_{i}
\end{align}
where $\lambda_{i}$ is the mean of unit $i$, $m_{0}$ is the intercept that follow a weakly informative Gaussian prior with mean zero and large variance, $\sum_{f = 1}^{F} \beta_f$ is the fixed effect of the covariates $X_{i f}$. Random effects such us splines for non linear effect of covariates $\boldsymbol{u_{i}}$ is included through the functions $\{\rho^r\}_{r=1}^{R}$, $\boldsymbol{b}$ is the spatial effects. \\
For the spatial effects, $\boldsymbol{b}$,  different spatial models for areal data can be assumed such as the Besag model \cite{besag1991bayesian}, or the extended Besag-York-Mollie model \cite{besag1991bayesian}, the Leroux model \cite{leroux2000estimation}, or the Dean's model \cite{dean2001detecting}. \\ \\
$ \pmb x = \{\eta_{ 1}, ...,\eta_{n},m_0, \beta_1, ..., \beta_{F}, \pmb\rho, \pmb b\}$ is called a latent field. 
with hyperparameters $\pmb\theta=\{\pmb\theta_{\pmb\rho}, \pmb\theta_{\pmb{b}}\}$, then the data $\pmb y$ is conditionally independent given the latent field and the hyperparameters such that the likelihood function is
 \begin{equation}
 \pi(\pmb x, \pmb\theta|\pmb y) = \prod_{i=1}^n f(y_i|x_i,\pmb\theta)\label{eq:lik0}
 \end{equation}

 \subsection{Prior specification and posterior propriety}
We assume prior independence amongst the parameters and as such we assign Gaussian priors to the latent field elements and various other prior to the hyperparameters as set out next.\\
For the latent field elements assume the following:
\begin{eqnarray}
&m_0 &\sim N(0,\tau_m^{-1}),\quad 
\pmb\beta|\tau_\beta \sim N(\pmb 0, \tau_\beta^{-1}\pmb I)\notag, \quad \\
& \pmb\rho|\pmb\theta_\rho & \sim N(\pmb 0, \pmb Q^{-1}_\rho),\quad \pmb b|\pmb\theta_{b} \sim N(\pmb 0, \pmb Q^{-1}_{b})\quad
\label{eq:priors0}
\end{eqnarray}
so that the joint prior for this part of the latent field is
\begin{equation*}
    \pmb x \sim N(\pmb 0, \pmb Q^{-1})
\end{equation*}
 where $\pmb Q^{-1}$ has a block diagonal structure as formed from \eqref{eq:priors0}.\\
 The vector of hyperparameters, $\pmb\theta$ is assigned a joint prior $\pi(\pmb\theta)$ which is composed of independent marginal proper priors of any shape (not necessarily Gaussian).\\ \\
The joint posterior of the unknown parameters, $\pmb x$ and $\pmb\theta$ from \eqref{eq:lik0} and \eqref{eq:priors0} is
 \begin{equation*}
     \pi(\pmb x,\pmb\theta|\pmb y)  \propto  \pi(\pmb y|\pmb x, \pmb\theta)\pi(\pmb x|\pmb\theta)\pi(\pmb\theta),
 \end{equation*}
 and based on the prior structures the posterior propriety holds.
 
\section{Quantile Regression}
Quantile regression describes the conditional quantile of the response variable given the explanatory variables, instead of the conditional mean. Let $Y$ be a real valued random variable. The $\alpha^{\text{th}}$ quantile of $Y$ is given by 
\begin{equation*}
Q(\alpha)=F^{-1}(\alpha)=\inf \{y: F(y) \geq \alpha\} \quad \text { for } \quad 0 \leq \alpha \leq 1,
\end{equation*}
where $F(y) = P(Y \leq y)$ is the Cumulative Distribution Function (CDF) of the random variable $Y$. Like the mean regression, a loss function is used in order to infer the parameters. The loss function of the quantile regression is the check loss function. Given that $0 \leq \alpha \leq 1$, $\forall x \in \mathbb{R}$ the quantile loss function is defined as
\begin{equation*}
\rho_{\alpha}(x)=\left\{\begin{array}{ll}
x \alpha & x \geq 0 \\
x(\alpha-1) & x<0
\end{array}\right.
\end{equation*}
An estimate of the $\alpha^{\text{th}}$ quantile of the random variable $Y$ can be obtained by minimizing the following risk function:
\begin{equation}\label{mini}
\underset{q_{\alpha} \in \mathbb{R}}{\operatorname{minimize}} \, \mathbb{E}\left[\rho_{\alpha}\left(Y-q_{\alpha}\right)\right]
\end{equation}
When $q_{\alpha}$ depends on the explanatory variables $\textbf{X}$, then it is called a conditional quantile. The estimate of the conditional quantile is called quantile regression. The quantile regression summarizes the relationship between $\textbf{X}$ and the quantile of Y. The estimate of the quantile regression can be written as 
\begin{equation} \label{opt}
\hat{q}_{\alpha} =\underset{q_{\alpha} \in \mathbb{R}}{\operatorname{argmin}} \, \mathbb{E}\left[\rho_{\alpha}\left(Y-q_{\alpha}\right)\right]
\end{equation}
Then $F(\hat{q}_{\alpha}) = \alpha$, where $F(y)$ is the CDF of the random variable $Y$.

\subsection{Model-based Quantile Regression}
The goal of the statistical analysis based on the 
Bayesian methods is to make inference from the posterior distribution for unknown parameters. Model-based quantile regression is an approach for quantile regression that considers the quantiles of the generating distribution proposed by \cite{padellini2018model}. This approach extends the Generalized Linear Mixed Model (GLMM) framework from modelling means to modelling quantiles. Two steps can do this extension. The first step is modeling the quantile; in this step, the quantile of the distribution is linked to the linear predictor through an invertible function $g$. The second step is mapping the quantile; in this step, the quantile is mapped to the parameter of the distribution through a map function $h$.
This approach can be applied to both frequentist and Bayesian frameworks. The resulting parameters of the Bayesian framework are all identifiable, making model-based quantile regression appealing in the Bayesian inference. To see these steps, let $F(y_{i} ; \lambda_{i})$ be the distribution of $Y_{i} | X_{i}$, where $\lambda_{i}$ is the parameter of the distribution. Given $0 	\leq \alpha 	\leq 1$, the $\alpha^{th}$ quantile of $Y_{i} | X_{i}$ is $q_{i,\alpha} = {Q}_{\alpha} (Y_{i} | X_{i})$.
The two steps can be written as follows:\\ \\
$\textbf{Step 1 - Modelling}.$ \\
The quantile $q_{i,\alpha}$ of the distribution $F(y_{i}, \lambda_{i})$ is modeled as follows:
\begin{equation*}
q_{i,\alpha}=g\left(\eta_{i,\alpha}\right), 
\end{equation*}
where $g$ is an invertible function, and $\eta_{i}^{\alpha}$ is the linear predictor for the level $\alpha$ quantile for $i=1, \ldots, n$. The linear predictor can include fixed effects, random effects, or both. Moreover, parametric or semi-parametric models can be included in this approach in order to study the impact of the covariates at different levels of the distribution and non-parametric models can be used for prediction.\\ \\
$\textbf{Step 2 - Mapping}$.\\
The quantile $q_{i,\alpha}$ is mapped to the parameter $\lambda_{i}$ of the distribution $F(y_{i} ; \lambda_{i})$ as 
\begin{equation} \label{tt}
\lambda_{i}=h\left(q_{i,\alpha}\right),
\end{equation}
where $h$ is an invertible map function. The map $h$ can be obtained by two steps. First, taking the inverse of the CDF ($F(y_{i} ; \lambda_{i})$) which give you the quantile function $Q(\alpha,\lambda_{i})$. Then, we write the parameter $\lambda_{i}$ as a function of the quantile, and that function is the map $h$. In this approach, the parameter $\lambda_{i}$ is modeled indirectly by the link between the quantiles of the generating distribution and $\lambda_{i}$. 
Unlike mean regression, when the parameter of the generating distribution links to the linear predictor through a function $\lambda_{i}=g(\eta_{i})$, in model-based quantile regression the parameter of the generating distribution is linked to the linear predictor through a composition function $\lambda_{i}=h(g(\eta_{i}))$. In other words, in the mean regression, the (GLMM) have a link function $\lambda_{i} = g(\eta_{i})$ to link the parameter of the generating distribution to the linear predictor. 

\subsection{Model-based quantile regression for count data}\label{sec:qr_count}
\begin{figure}[H]
\begin{center}
\includegraphics[scale=0.8]{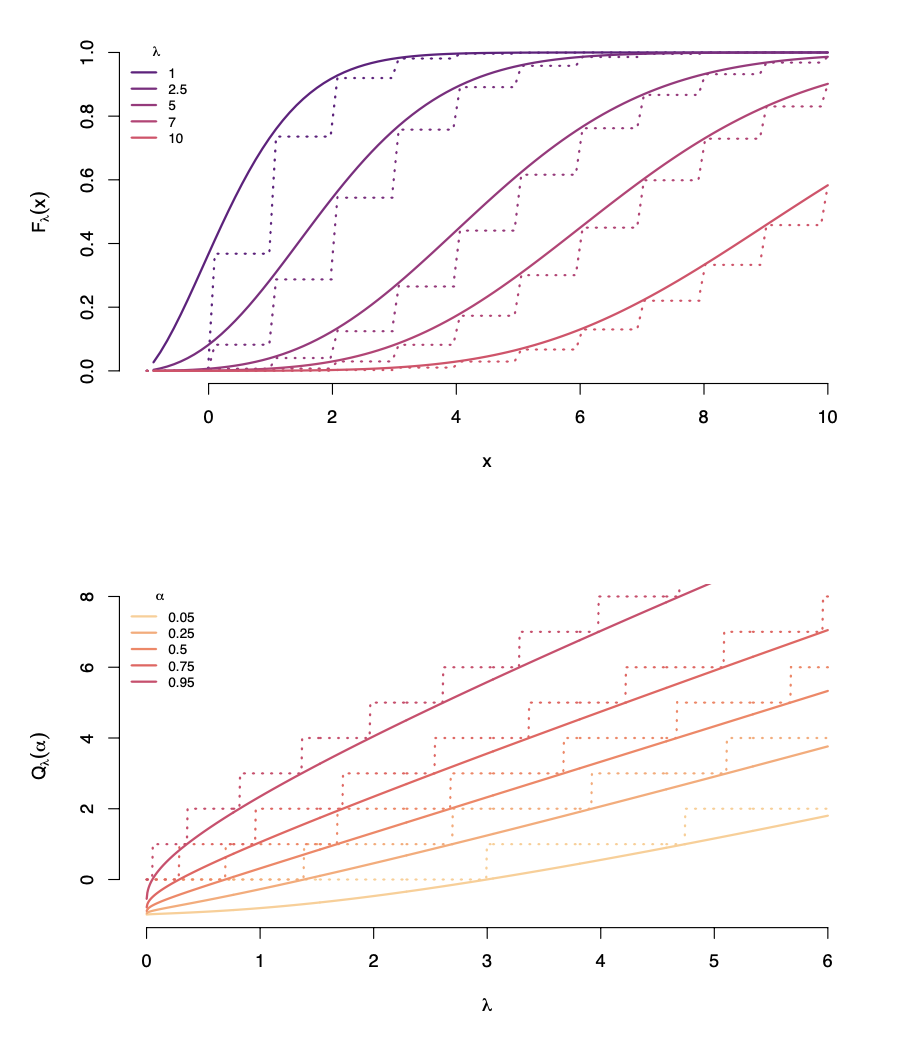}
\caption{(Top) The CDF of the discrete Poisson
(dashed line), and the CDF of the continuous Poisson (solid line). (Bottom) Quantile function of the discrete (dashed line) and continuous (continuous line) Poisson distributions}
\label{tu1}
\cite{padellini2018model}
\end{center}
\end{figure}

The extension of model-based quantile regression for discrete random variables is not straight-forward since the objective function in \eqref{opt} is non-differentiable for discrete random variables. The positive mass of the points for the discrete variable prevent the sample quantile from having an asymptotic distribution. Additionally, it is not easy to apply the modeling and mapping steps of model-based quantile regression to discrete data. First, in the modeling step, the common models for $g$ are the log for count data and the logit for binary data, and they are continuous functions. Therefore, the model  $q_{i,\alpha}=g\left(\eta_{i,\alpha}\right)$ is not appropriate, since the quantile which is on the left hand side is discrete whereas the function $g$ is continuous. The second reason, in the mapping step, it is hard to get the map $h$ because the CDF of the discrete is non-invertible, which implies that there is no unique $\lambda_{i}$ to generate 
each quantile, as one can be seen in Figure \ref{tu1}.\\ \\
To address these issues, \cite{padellini2018model} approximated discrete distributions by continuous counterparts, and then model the quantile for the continuous version instead of the discrete. The continuous counterpart is obtained by interpolating the cumulative distribution function (CDF) of the discrete random variable. The model-based quantile method can be applied to discrete variables if their CDF can be expressed as 
\begin{equation*}
F_{X}(x ; \lambda)=\mathbb{P}(X \leq x)=k(\lfloor x\rfloor, \lambda)
\end{equation*}
where $k$ is a continuous function, and $X$ is a discrete random variable. The interpolation can be obtained by removing the floor operator, so that $k(x,\lambda)$ is the CDF of the continuous version of $X$, assigned $X^{\prime}$. By definition of the floor, for all integers $x$ 
\begin{equation*}
F_{X}(x)=k(\lfloor x\rfloor, \lambda)=k(x, \lambda)=F_{X^{\prime}}(x).
\end{equation*}
The continuous distribution of $X'$ is considered as a continuous generalization of the original variable because the two CDFs are equal for all integer values $x$.
The advantage of working with the continuous version of a discrete distribution in the Bayesian framework is that a likelihood function can be obtained by using the model-based quantile method, since the sample quantiles for a discrete random variable are generally not asymptotically normal \cite{padellini2018model}.

\subsubsection{Continuous Poisson}
Here we present the details on the approximation of the discrete Poisson distribution with a continuous Poisson counterpart.\\ \\
The CDF of a Poisson distribution can be expressed as the ratio of an incomplete and regular Gamma function as follows:
\begin{equation}
X \sim \operatorname{Poisson}(\lambda) \quad F_{X}(x)=\mathbb{P}(X \leq x)=\frac{\Gamma(\lfloor x\rfloor+1, \lambda)}{\Gamma(\lfloor x\rfloor+1)} \quad x \geq 0
\label{eq:discrete_poi}
\end{equation}
where $\Gamma(x, \lambda)=\int_{\lambda}^{\infty} e^{-s} s^{x-1} \mathrm{~d} s$ is the upper incomplete Gamma function. Following Section \ref{sec:qr_count}, the Continuous Poisson is then defined from \eqref{eq:discrete_poi} as
\begin{equation*}
X^{\prime} \sim \text { Continuous Poisson }(\lambda) \quad F_{X^{\prime}}(x)=\mathbb{P}\left(X^{\prime} \leq x\right)=\frac{\Gamma(x+1, \lambda)}{\Gamma(x+1)} \quad x>-1
\end{equation*}
The reason for changing the support from $x\geq 0$ to $x>-1$ is to avoid mass at $0$, so there will be no jump on the CDF of the Continuous Poisson (CP)as illustrated in Figure \ref{tu1}. If the support remains the same, then the value of the CP will be $0$ if $ x<0$ and about $0.4$ at $x=0$, which introduces a jump at zero. However, if the support is $x >-1$, then there will be no jump at zero because the CDF of the CP will be $0$ at $x=-1$, then an interpolation
will be applied from $x=-1$ to $x=0$. The
Continuous and discrete Poisson random variables can be related as $X=\left\lceil X^{\prime}\right\rceil$. \\ \\
The model-based quantile regression model for Poisson data is then defined for $Y_{i} \mid \eta_{i}$ a continuous Poisson random variable with parameter $\lambda_i$ as  
\begin{eqnarray} \label{paula}
q_{i,\alpha}&=&g\left(\eta_{i,\alpha}\right)=\exp \left\{\eta_{i,\alpha}\right\} \notag\\
\lambda_{i}&=&h\left(q_{i,\alpha}\right)=\frac{\Gamma^{-1}\left(q_{i,\alpha}+1,1-\alpha\right)}{\Gamma\left(q_{i,\alpha}+1\right)}.
\end{eqnarray}

\subsection{Model-based quantile regression for disease mapping}\label{sec:qr_count}
From Sections \ref{sec:dis_map} and \ref{sec:qr_count} we can define a model-based quantile regression model for disease mapping. One issue that remains is how to decompose the expected number of cases into the local expectation, $E_i$ and the relative risk $\lambda_i$. In the case of modeling the quantile instead of the mean there are two options as discussed by \cite{padellini2018model}:
\begin{itemize}
    \item Include $E_{i}$ in the linear model as an offset 
    \begin{eqnarray}
q_{i,\alpha}&=&\exp \left\{\eta_{i,\alpha}+\log \left(E_{i}\right)\right\}=E_{i} \exp \left\{\eta_{i,\alpha}\right\} \notag\\
\lambda_{i}&=&\frac{\Gamma^{-1}\left(q_{i,\alpha}+1,1-\alpha\right)}{\Gamma\left(q_{i,\alpha}+1\right)}
\label{eq:qs_model}
\end{eqnarray}

\item  Consider it as a scaling of the parameter of the distribution 

 \begin{eqnarray}
q_{i,\alpha} &=&\exp \left\{\eta_{i,\alpha}\right\} \notag \\
\lambda_{i} &=&E_{i} \frac{\Gamma^{-1}\left(q_{i,\alpha}+1,1-\alpha\right)}{\Gamma\left(q_{i,\alpha}+1\right)}
\label{eq:global_model}
\end{eqnarray}
\end{itemize}
 These
  two approaches are equivalent in the Poisson mean regression, but not equal in the Poisson quantile regression and the choice of approach depends on the purpose of the analysis. If the focus of the study is to infer a quantile-specific model then \eqref{eq:qs_model} is more appropriate whereas \eqref{eq:global_model} can be considered as a model for the parameter $\lambda_i$. 

\section{Bayesian Joint Quantile Disease Mapping} \label{jointq}

The main goal of disease mapping is to estimate the relative risk of diseases across regions. Sometimes specific diseases have similar spatial patterns due to sharing the same etiologies. In this case, these diseases have some dependence, and it would be more appropriate to model them jointly rather than separately. Moreover, sometimes the dependence might be in different quantiles between the diseases or some diseases could inhibit the occurence of another disease. The proposed joint quantile disease mapping model links different quantiles of multiple diseases by a more general framework by considering dependence not in the mean, but in the quantiles.
\subsection{Model specification}
The joint quantile model for two diseases can be formulated as:

\begin{eqnarray}
y_{i 1} &\sim& \text { Poisson }\left(\lambda_{i 1}\right) \notag\\
y_{i 2} &\sim& \text { Poisson }\left(\lambda_{i 2}\right) \notag\\
\log (q_{i 1, \alpha_{1}}) &=& \log(\eta_{i 1,\alpha_{1}}) =  m_{1} + \sum_{f = 1}^{F_1} \beta_f X_{i f} + \sum_{r=1}^{R_1} \rho^r(u_{i r}) + b_{i 1} + S_{i} \label{qa} \\
\log (q_{i 2,\alpha_{2}}) &=& \log(\eta_{i 2,\alpha_{2}}) = m_{2} + \sum_{f = 1}^{F_2} \gamma_f Z_{i f} + \sum_{r=1}^{R_2} \xi^r(v_{i r})+ b_{i 2}  + c \, S_{i}\label{qa2}
\end{eqnarray}
where $\lambda_{i k}$ is the mean of unit $i$ for disease $k$ and it is mapped to the $\alpha_{k}$ level quantile $q_{i k, \alpha_{k}}$ as in \eqref{paula}. In the modeling part, $m_{k}$ is a disease-specific intercept that follows a weakly informative Gaussian prior with zero mean and large variance, $b_{i k}$ is a spatial random effect and $S_{i}$ is the shared spatial component. The model also incorporates fixed effects of covariates $\pmb X_i$ and $\pmb Z_i$ by $\sum_{f = 1}^{F_1} \beta_f X_{i f}$ in \eqref{qa} and $\sum_{f = 1}^{F_2} \gamma_f Z_{i f}$ in \eqref{qa2}, respectively. Various random effects such as splines for non-linear effects of covariates $\pmb u_i$ and $\pmb v_i$  is included through functions $\{\rho^r\}_{r=1}^{R_1}$ in \eqref{qa} and $\{\xi^r\}_{r=1}^{R_2}$ in \eqref{qa2}, respectively.  \\ \\
For the disease-specific spatial effects, $\pmb b_{k}$, various spatial models for areal data can be assumed such as the Besag model \cite{besag1991bayesian}, or the extended Besag-York-Mollie model \cite{besag1991bayesian}, the Leroux model \cite{leroux2000estimation}, or the Dean's model \cite{dean2001detecting}.\\
The shared spatial component, $\pmb S$, that links the two diseases through their quantiles is assumed to be a Besag effects with precision matrix 
 $Q=\left(Q_{i j}\right)$, where for $j \neq i$, $Q_{i i}=\tau\left(n_{i}+d\right)$, $Q_{i j}=-\tau$, and $n_{i}$ is the number of neighbours of node $i$. 
 The parameter $c \in \Re$ is used to scale the shared component and correlate the two diseases in space. \\ \\
 Now we can collect $m_1, m_2, \pmb\beta, \pmb\rho, \pmb b_1, \pmb S, \pmb\gamma, \pmb\xi, \pmb b_2$ together with the linear predictors $\eta_{1 1, \alpha_1}, ...,\eta_{n 1, \alpha_1},\eta_{1 2, \alpha_2}, ...,\eta_{n 2, \alpha_2}$ and form the latent field 
 \begin{equation*}
     \pmb x = \{\eta_{1 1, \alpha_1}, ...,\eta_{n 1, \alpha_1},\eta_{1 2, \alpha_2}, ...,\eta_{n 2, \alpha_2},m_1, m_2, \beta_1, ..., \beta_{F_1}, \pmb\rho, \pmb b_1, \pmb S, \pmb\gamma, \pmb\xi, \pmb b_2\}
 \end{equation*}
 with hyperparameters $\pmb\theta=\{c, \tau, d, \pmb\theta_{\pmb\rho}, \pmb\theta_{\pmb\xi}, \pmb\theta_{\pmb{b}_1}, \pmb\theta_{\pmb{b}_2}\}$, then we have that the data $\pmb y = \{\pmb y_1, \pmb y_2, \}$ is conditionally independent given the latent field and the hyperparameters such that the likelihood function is
 \begin{equation}
 \pi(\pmb x, \pmb\theta|\pmb y) = \prod_{i=1}^n f(y_i|x_i,\pmb\theta)\label{eq:lik}
 \end{equation}

\subsection{Prior specification and posterior propriety}
We assume prior independence amongst the parameters and as such we assign Gaussian priors to the latent field elements and various other prior to the hyperparameters as set out next.\\
For the latent field elements assume the following:\\
\begin{eqnarray}
m_k &\sim& N(0,\tau_m^{-1}),\quad
\pmb\beta|\tau_\beta \sim N(\pmb 0, \tau_\beta^{-1}\pmb I),\quad \pmb\gamma|\tau_\gamma \sim N(\pmb 0, \tau_\gamma^{-1}\pmb I)\notag\\
\pmb\rho|\pmb\theta_\rho &\sim& N(\pmb 0, \pmb Q^{-1}_\rho),\quad
\pmb\xi|\pmb\theta_\xi \sim N(\pmb 0, \pmb Q^{-1}_\xi),\quad \pmb S|\tau, d \sim N(\pmb 0, \pmb Q^{-1}_S)\notag\\
\pmb b_1|\pmb\theta_{b_1} &\sim& N(\pmb 0, \pmb Q^{-1}_{b_1}),\quad
\pmb b_2|\pmb\theta_{b_2} \sim N(\pmb 0, \pmb Q^{-1}_{b_2})
\label{eq:priors}
\end{eqnarray}
so that the joint prior for these elements of the latent field is
\begin{equation*}
    \pmb x \sim N(\pmb 0, \pmb Q^{-1})
\end{equation*}
 where $\pmb Q^{-1}$ has a block diagonal structure as formed from \eqref{eq:priors}.\\
 The vector of hyperparameters, $\pmb\theta$ is assigned a joint prior $\pi(\pmb\theta)$ which is composed of independent marginal proper priors of any shape (not necessarily Gaussian).\\ \\
 The shared spatial field is assumed to follow Besag model but with an additional parameter $d$, to ensure a proper prior of $\pmb S$ as follows:
\begin{eqnarray*}
    \pmb S|\tau, d &\sim& N(\pmb 0, \pmb Q^{-1}_S)
\end{eqnarray*}with the entries of $ \pmb Q_S$ as follows:

\begin{equation}
Q_{S,ii} = \tau(n_i+d)\quad \text{and}\quad Q_{S,ij} = -\tau,  \label{eq:besagproper}
\end{equation}
for $i\neq j$, and $j$ is in the neighbourhood of $i$.\\ \\
The disease-specific spatial fields $\pmb b_1$ and $\pmb b_2$ are assumed to follow uncorrelated BYM/CAR models where we reparameterize the precision matrix similar to \cite{banerjee2014} to have more orthogonal parameters resulting in useful practical interpretation of the weight parameter, $\phi$. One issue with the proper CAR parameterization proposed by \cite{banerjee2014} is that the weight parameter is still not practically a weight since the unstructered effect and the Besag field might have different generalized variances. To alleviate this issue, we scale both the unstructured and Besag components to have the same geometric mean and define the proper scaled BYM field as
\begin{equation}
    \pmb b_k = \frac{1}{\sqrt{\tau_{b_k}}}\left(\sqrt{1-\phi_{b_k}}\pmb b_k^1 + \sqrt{\phi_{b_k}}\pmb b_k^2\right)
\end{equation}
with $\pmb b_k^1$ a scaled IID effect and $\pmb b_k^2$ a scaled Besag effect as in \eqref{eq:besagproper}. With this formulation, $\phi_{b_k}$ can be interpreted as the proportion of the marginal variance explained by the spatial effect, and $1-\phi_{b_k}$ is the proportion of the marginal variance explained by the unstructured effect.\\ \\

 The joint posterior of the unknown parameters, $\pmb x$ and $\pmb\theta$ from \eqref{eq:lik} and \eqref{eq:priors} is
 \begin{equation*}
     \pi(\pmb x,\pmb\theta|\pmb y)  \propto  \pi(\pmb y|\pmb x, \pmb\theta)\pi(\pmb x|\pmb\theta)\pi(\pmb\theta),
 \end{equation*}
 and based on the prior structures the posterior propriety holds.
 \subsection{Approximate inference using INLA}
Computational Bayesian inference can be achieved largely in one of two ways, either through sampling-based methods like Markov Chain Monte Carlo (MCMC) and deviants or approximately using approximate methods like Variational methods or Laplace approximations like Integrated Nested Laplace Approximation (INLA). INLA, as introduced by \cite{rue2009approximate}, has been shown to be widely applicable to various statistical models; in particular, to the latent Gaussian models class of which disease mapping models are included \cite{martinez2015climatic,moraga2019geospatial,moraga2021bayesian,ugarte2014fitting}\\ \\
INLA employs a series of Laplace approximations and numerical integration to perform approximate Bayesian inference through numerically approximating the posterior densities of the latent field and hyperparameters. For data $\pmb y$, latent field $\pmb x$ and hyperparameters $\pmb \theta$, INLA can be summarized as follows:
\begin{enumerate}
\item Find the $m$-variate Gaussian approximation of $\pi(\pmb x|\pmb\theta, \pmb y)$ at the mode $\pmb\mu(\pmb\theta)$ with matching curvature using the Hessian of $\pi(\pmb x|\pmb\theta, \pmb y)$ at the mode $\pmb\mu(\pmb\theta)$.
\item Let 
\begin{equation}
\tilde{\pi}(\pmb\theta|\pmb y) \propto \frac{\pi(\pmb x^*,\pmb\theta| \pmb y)}{\pi_G(\pmb x^*|\pmb\theta, \pmb y)} |_{\pmb x^* = \pmb\mu(\pmb\theta)}
\end{equation} 
and locate the mode of $\tilde{\pi}(\pmb\theta|\pmb y)$ and find a set of integration points $\pmb\theta_k, k = 1,2,...,T$ in the area of the highest probability mass.
\item Calculate 
\begin{equation}
\tilde{\pi}(\theta_j|\pmb y) = \int_{\pmb\theta_{-j}}\tilde{\pi}(\pmb\theta|\pmb y)d\pmb\theta_{-j}
\end{equation}
where we note that this is a low-dimensional integral since $p$ is generally small.
\item Now define 
\begin{eqnarray}
\tilde{\pi}(x_i|\pmb\theta_k, \pmb y) \approx \frac{\pi(\pmb x^*,\pmb\theta_k| \pmb y)}{\pi_G(\pmb x^*_{-i}|x_i, \pmb\theta_k, \pmb y)} |_{\pmb x^*_{-i} = \pmb\mu_{-i}(\pmb\theta_k)}
\end{eqnarray}
with $\pi_G(\pmb x^*_{-i}|x_i, \pmb\theta, \pmb y)$ the $(m-1)$-variate Gaussian approximation at the mode $\pmb\mu_{-i}(\pmb\theta)$ for the $T$ configuration points $\pmb\theta_k, k = 1,2,...,T$, and calculate
\begin{equation}
\tilde{\pi}(x_i|\pmb y) \approx \sum_{k=1}^T\tilde{\pi}(x_i|\pmb\theta_k, \pmb y)\tilde{\pi}(\theta_k|\pmb y)\Delta_k
\end{equation} where $\tilde{\pi}(\theta_k|\pmb y)$ is from step 3, with $\Delta_k$ the step size.
\end{enumerate}
Various simplifications to the approximations have been proposed as well in order to achieve increased computational efficiency such as an empirical Bayes approach where the integration points $\pmb\theta_k$ are all set to the mode of $\tilde{\pi}(\pmb\theta|\pmb y)$, which is named a \textit{Simplified Laplace approximation strategy}.

\subsection{Simulation Study}
\begin{figure}[H]
\begin{center}
\hspace*
{-.2cm}\includegraphics[width = 6cm]{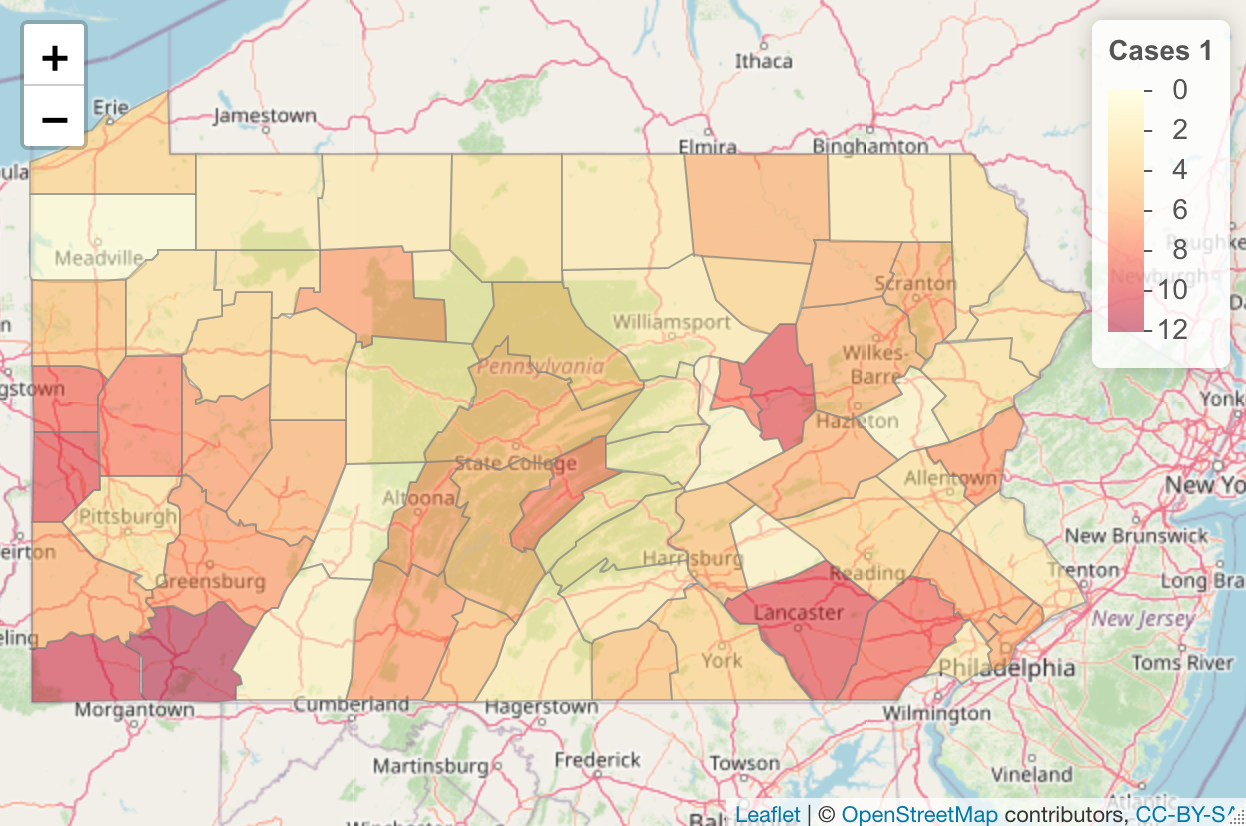}
\includegraphics[width = 6cm]{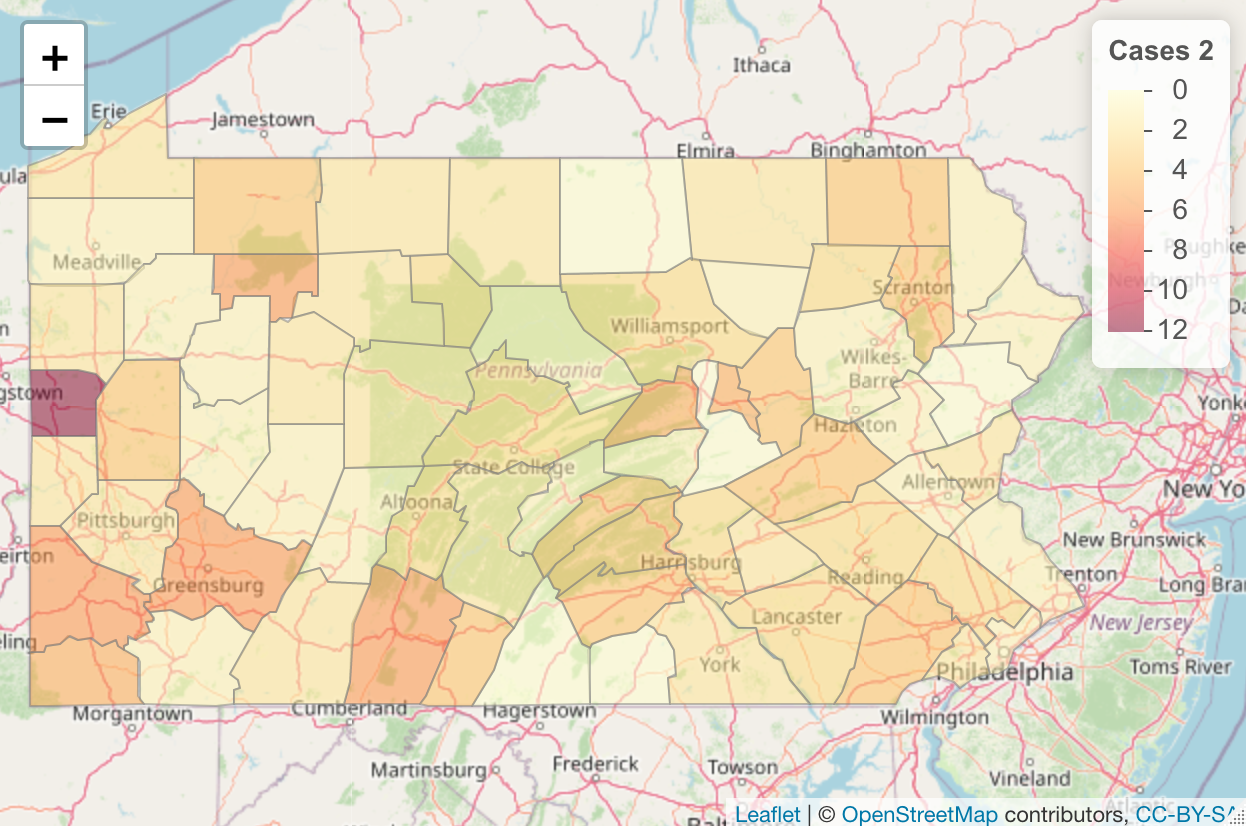}
\caption{
One realization of cases of disease 1 (left) and disease 2 (right).}
\label{simulated cases}
\end{center}
\end{figure}

The code for this simulation study is available at \url{https://github.com/JanetVN1201/Code_for_papers/tree/main/Joint\%20quantile\%20disease\%20mapping\%20}.\\ \\
In this part, simulated independent and correlated data was added to the Pennsylvania map, which is considered as a connected graph of size $67$. Figure \ref{simulated cases} shows one realization of the correlated data that was added to the Pennsylvania map.

\begin{figure}[H]
\begin{center}
\includegraphics[scale=.55]{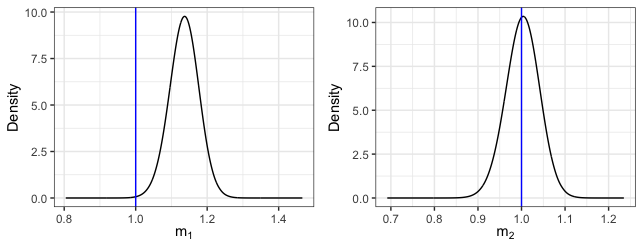}
\caption{The posterior distribution for the intercepts}
\label{jointq2}
\end{center}
\end{figure}

\begin{figure}
  \includegraphics[width = 5.2cm]{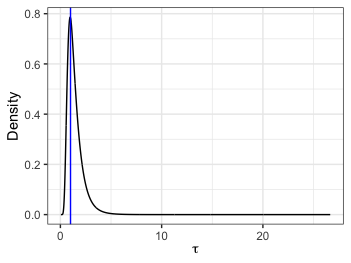}  
\includegraphics[width = 5.2cm]{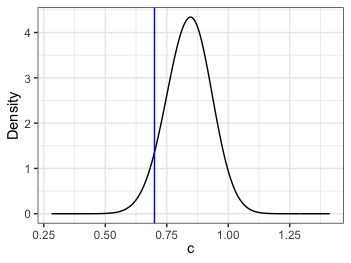}  
  \includegraphics[width = 5.2cm]{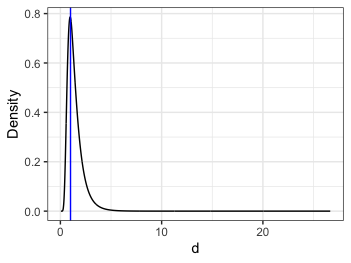}  
\caption{The posterior distribution for the hyperparameters}
\label{hyperqusim}
\end{figure}

The correlated data was generated as follows.
\begin{eqnarray*}
y_{i 1} &\sim& \text { Poisson }\left(\lambda_{i 1}\right)\quad \text{with} \quad  \log (q_{i 1, 0.2}) = 1 + S_{i}\\
y_{i 2} &\sim& \text { Poisson }\left( \lambda_{i 2}\right)\quad \text{with} \quad  \log (q_{i 2,0.8}) = 1 + 0.7\, S_{i}
\end{eqnarray*}
where $S_{i}$ is the shared component that follow a Besag proper model \eqref{eq:besagproper} with precision matrix $Q$, where
$Q_{i i}=\left(n_{i}+1\right)$ and $Q_{i j}=-1$ for $j \neq i$. In other words, the extra term added on the diagonal is $d= 1$, and the precision parameter $\tau = 1$. \\
There are two hyperparameters for this model: a parameter $d$ for controlling the properness, and a parameter $\tau > 0$ that is a scaling parameter. The hyperpriors defined on the $\log$ scale of these hyperparameters are as follows. 
\begin{eqnarray*}
\log(d) &\sim& \text{loggamma} (1,1)\\
\log(\tau) &\sim& \text{loggamma} (1,5e-04)
\end{eqnarray*}
The estimated values of the parameters obtained by R-INLA are similar to the true values as seen in Tables \ref{jointq1} and \ref{jointq3}, and Figures \ref{jointq2} and \ref{hyperqusim}.
\begin{table}[hbt!]
\centering
\begin{tabular}{ccccc}
\hline
\quad & \text{mean}  & \text{sd}  & \text{0.025quant} & \text{0.975quant}  \\
\hline
$m_1$ & 1.137  & 0.042  & 1.053 & 1.218  \\
$m_2$ & 1.003  & 0.039  & 0.923  & 1.079  \\
\hline
\end{tabular}
\caption{The estimated intercepts and the 95\% credible intervals.}
\label{jointq1}
\end{table}

\begin{table}[htb!]
\centering
\begin{tabular}{ccccc}
\hline
\quad & \text{mean}  & \text{0.025quant} & \text{0.975quant}  & \text{mode} \\
\hline
$\tau$ & 1.337  & 0.77  & 2.074 & 1.257  \\
$d$  & 1.409 & 0.472  & 3.612 & 0.887  \\
$c$  & 0.838  & 0.636  & 1.034  & 0.845 \\
\hline
\end{tabular}
\caption{The estimated hyperparameters and the 95\% credible intervals.}
\label{jointq3}
\end{table}

 Model selection criteria are presented in Tables \ref{jointq4} and \ref{jointq5} and it shows a preference for the joint quantile model when the data are correlated and a preference for the separate models, which are \eqref{qa} and \eqref{qa2} without the shared components, when the data are independent. This indicates stable estimation and the model's ability to distinctly estimate an associated joint model if needed.

\begin{table}[htb!]
\centering
\begin{tabular}{lcc}
\hline
\quad & \text{DIC}  & \text{WAIC} \\
\hline
\textbf{Separate 1} & 3311  & 3350  \\
\textbf{Separate 2} & 2439 & 2452  \\
\textbf{Sum of Separates} & 5750  & 5802 \\
\textbf{Joint quantile} & 5642 & 5660 \\
\hline
\end{tabular}
\caption{The DIC and WAIC for correlated data.}
\label{jointq4}
\end{table}

\begin{table}[htb!]
\centering
\begin{tabular}{lcc}
\hline
\quad & \text{DIC}  & \text{WAIC} \\
\hline
\textbf{Separate 1} & 2951  & 2967 \\
\textbf{Separate 2} & 2925 & 2953  \\
\textbf{Sum of Separates} & 5876  & 5919 \\
\textbf{Joint quantile} & 6002 & 6100 \\
\hline
\end{tabular}
\caption{The DIC and WAIC for independent data.}
\label{jointq5}
\end{table}

\section{Joint quantile disease mapping model for Malaria and G6PD}
In this section, we fit the Bayesian model by using R-INLA to estimate the risks of malaria and G6PD deficiency in some African countries by using separate and joint quantile models. The code for this analysis is available at \url{https://github.com/JanetVN1201/Code_for_papers/tree/main/Joint\%20quantile\%20disease\%20mapping\%20}.

\subsection{Exploratory data analysis} \label{data}
\begin{figure}[H]
\begin{center}
\hspace*
{-.2cm}\includegraphics[width = 7cm]{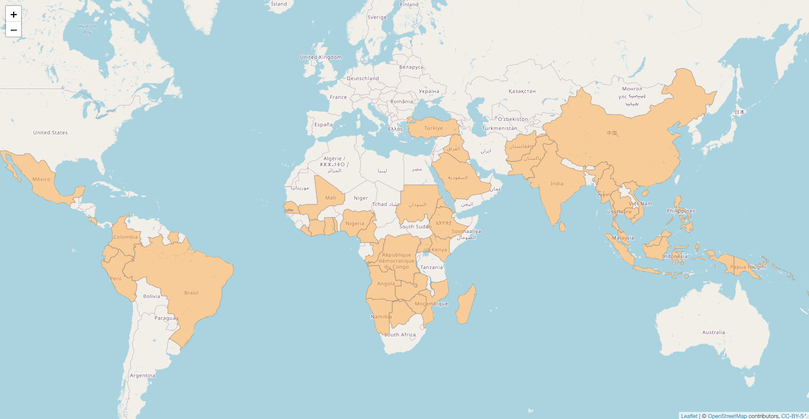}
\includegraphics[width = 7cm]{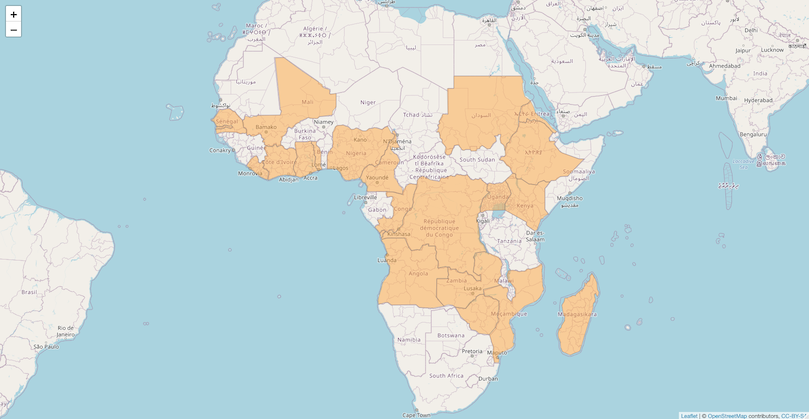}
\caption{
Countries where both G6PD deficiency and malaria cases exist (left) on the African continent (right)}
\label{common c}
\end{center}
\end{figure}

\begin{figure}
\begin{subfigure}{.3\textwidth}
  \includegraphics[width = 5cm]{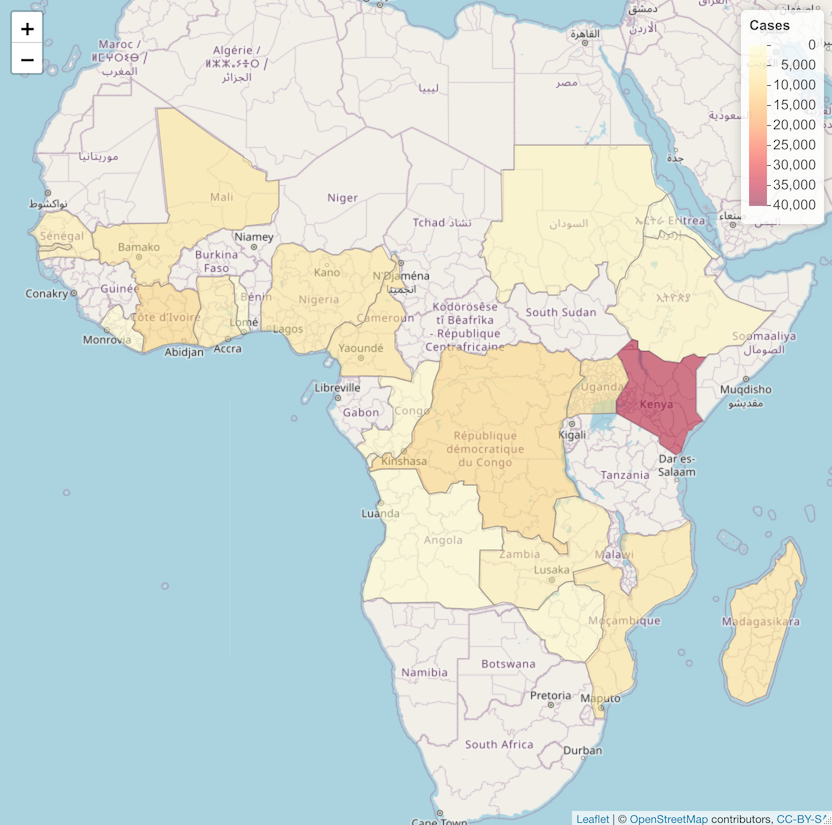}  
  \caption{Observed cases}
  \label{observed.m}
\end{subfigure}
\begin{subfigure}{.3\textwidth}
\includegraphics[width = 5cm]{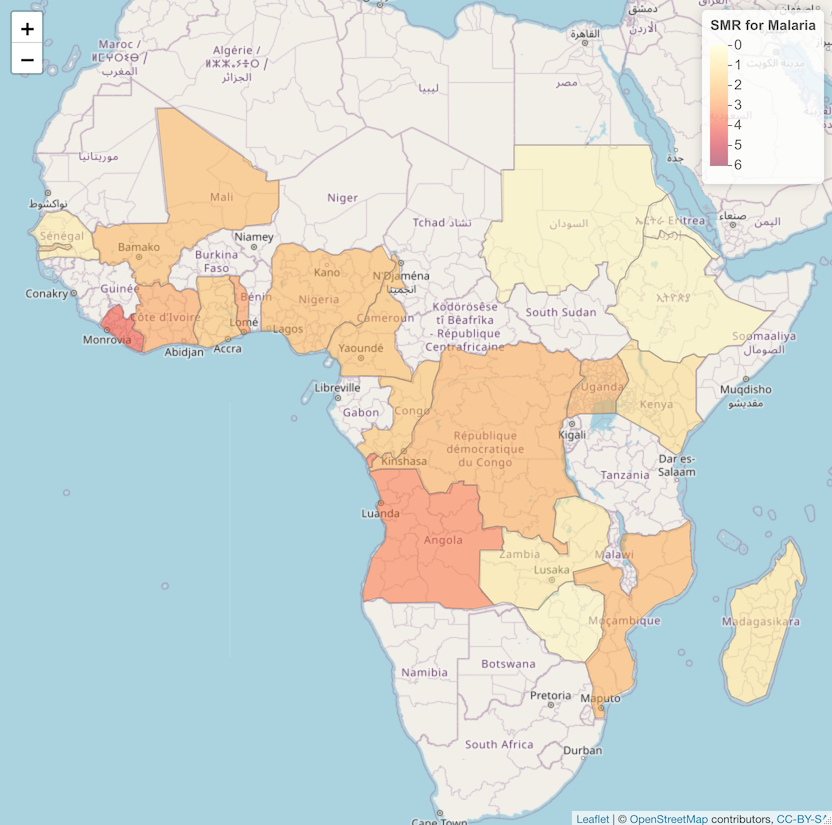}  
  \caption{SMR}
  \label{smr.m}
\end{subfigure}
\begin{subfigure}{.3\textwidth}
  \includegraphics[width = 5cm]{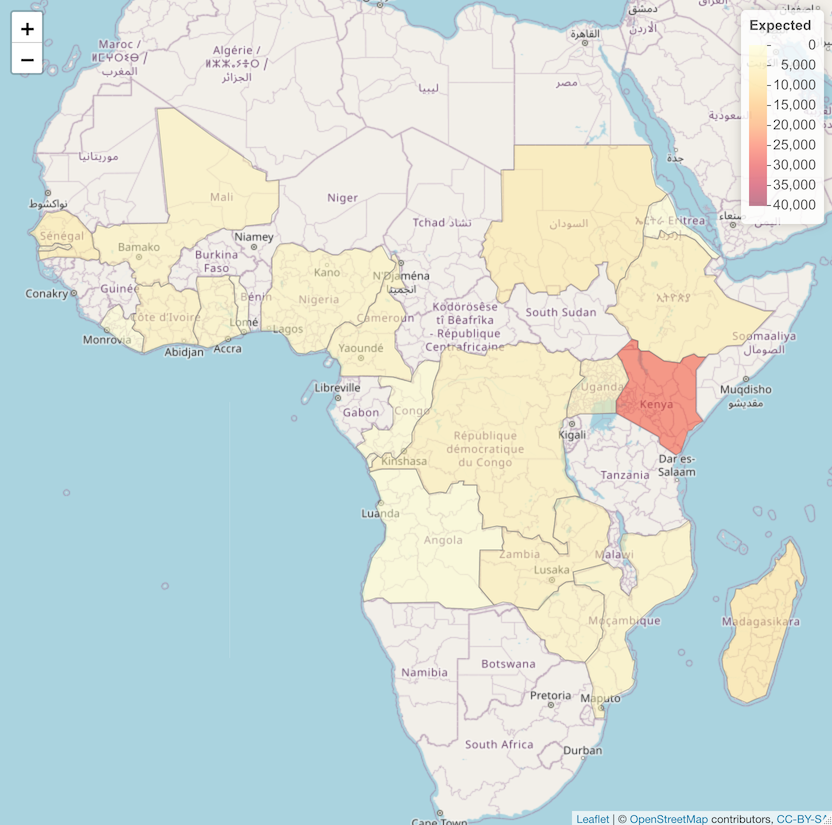}  
  \caption{Expected cases}
  \label{e.m}
\end{subfigure}

\begin{subfigure}{.3\textwidth}
  \includegraphics[width = 5cm]{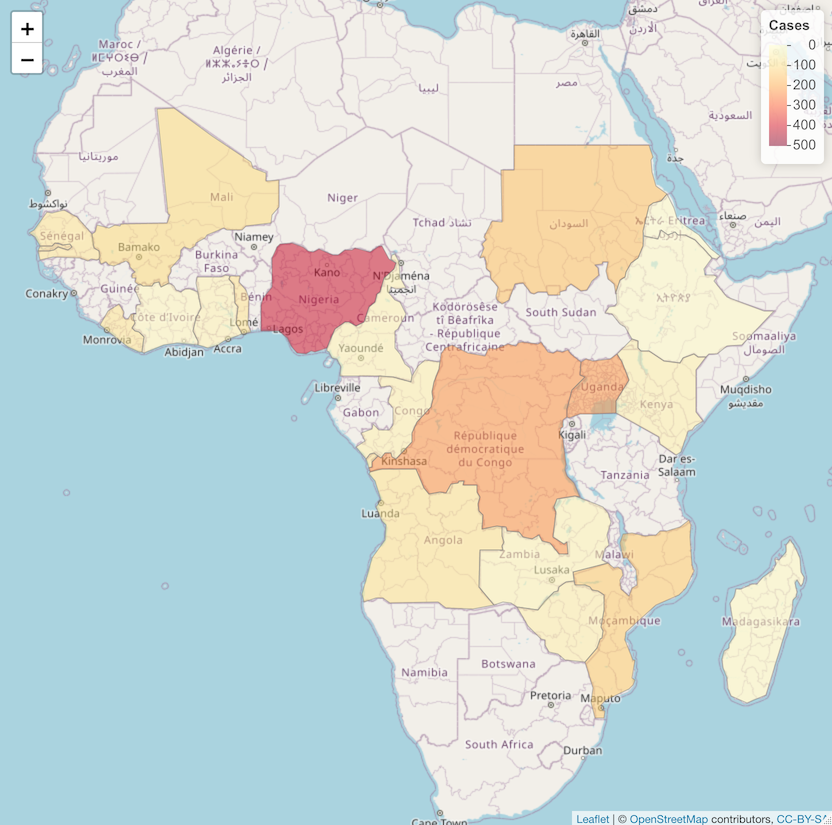}  
  \caption{Observed cases}
  \label{observed.d}
\end{subfigure}
\begin{subfigure}{.3\textwidth}
\includegraphics[width = 5cm]{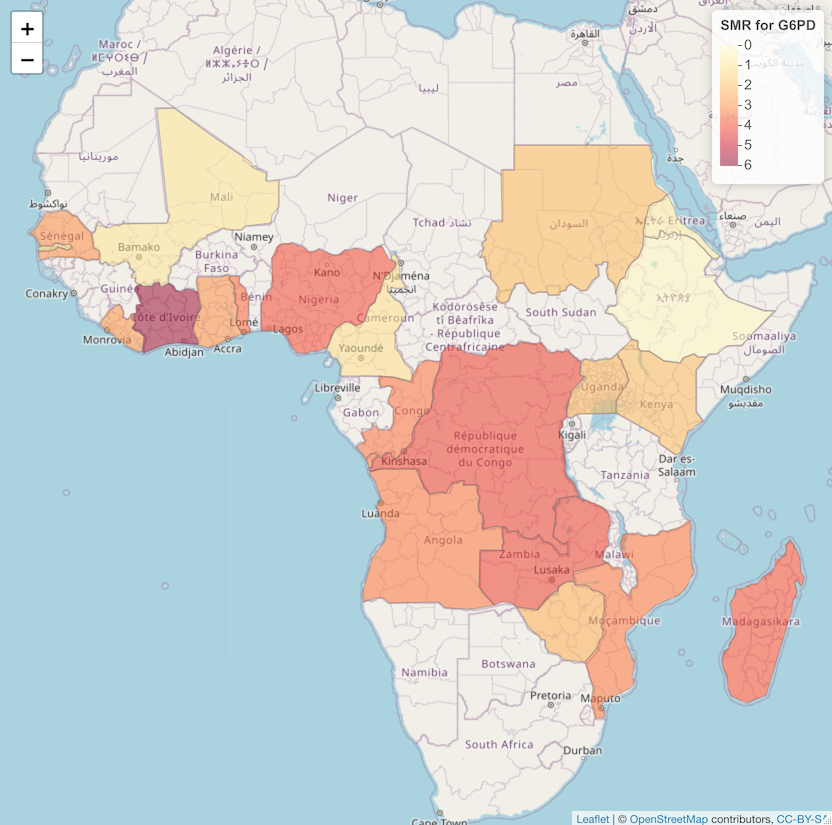}  
  \caption{SMR}
  \label{smr.d}
\end{subfigure}
\begin{subfigure}{.3\textwidth}
  \includegraphics[width = 5cm]{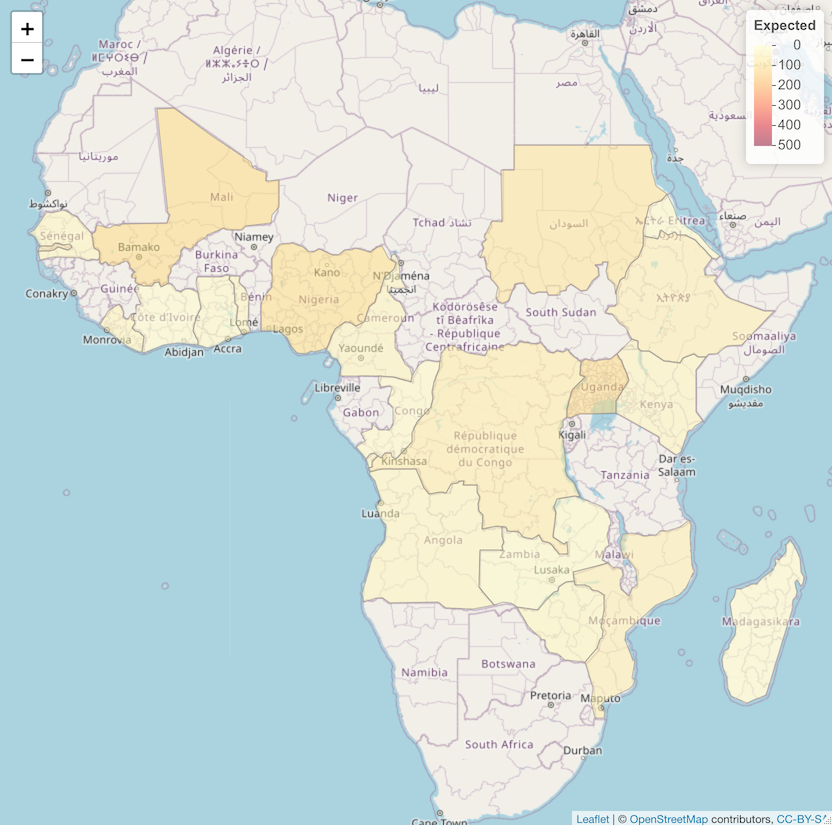}  
  \caption{Expected cases}
  \label{e.d}
\end{subfigure}
\caption{Summaries from exploratory data analysis for Malaria (top row) and G6PD (bottom row)}
\label{fig:fig}
\end{figure}

The malaria cases and G6PD cases per region is obtained from \url{https://malariaatlas.org/}. Various country-level covariates can be used in our model but for the motivating example the emphasis is placed on the joint component, even though for a thorough analysis of the data itself, various fixed and random effects might be considered. \\ \\
We selected only the countries for which information for both Malaria and G6PD is available, as indicated in Figure \ref{common c}. 
According to Figure \ref{common c}, the countries are distributed around the world. Since we want to investigate the spatial correlation we consider the African continent so that most countries have some neighbours as in Figure \ref{common c}.
In Figure \ref{fig:fig}, \ref{smr.m} and \ref{smr.d} the SMRs for malaria and G6PD deficiency. It can be seen that, in general, the risk of G6PD deficiency is higher than the risk of malaria because G6PD deficiency has a higher SMR. Some countries like Abidjan and Madagascar that considered to have the highest risk of G6PD deficiency, they have the lowest risk of malaria according to the SMR values, which could indicate a prohibitive relationship between these two diseases. 
The observed cases for malaria and G6PD can be seen in Figures \ref{observed.m} and \ref{observed.d} respectively. Kenya has the highest number of malaria cases, while Nigeria has the highest number of G6PD deficiency cases. 

\subsection{Results}
It is believed that G6PD deficiency limits the occurrence of malaria \cite{beutler1994G6PD,allison1961malaria}. We expect a correlation between a high quantile of G6PD and a low quantile of malaria. Therefore we applied the joint quantile model that was discussed in section \ref{jointq} where $y_{i1}$ and $y_{i2}$ represent the cases of malaria and G6PD deficiency, respectively. The quantile levels are $\alpha_{1} = 0.2$ and $\alpha_{2} = 0.8$.\\ \\
Table \ref{apjq0} shows the overall means for the malaria and the G6PD deficiency. Table \ref{apjq1} shows the estimation of the hyperparameters. The precision of the random effects indicate that most of the spatial variability comes from the shared component. There is a significant correlation between a high quantile of G6PD deficiency and a low quantile of malaria as observed from the point estimate and the credible interval of $c$, which is the coefficient of the shared component. This finding is consistent with the studies \cite{beutler1994G6PD,allison1961malaria}.\\ \\
In contrast, Table \ref{opp} shows the estimation of the hyperparameters for a high quantile of malaria with a low quantile of G6PD deficiency. As can be seen based on the credible interval of $c$, there is no significant correlation between these two quantiles. This is expected because having G6PD deficiency protects you from having malaria, but malaria does not influence G6PD deficiency.\\ \\
Figure \ref{jointspatialqant} presents maps of the spatial effects. Figures \ref{sharedq} and \ref{sharedq1} show the shared spatial effect for malaria and G6PD deficiency, respectively. The structure of the effects appear similar. However, the shared spatial effect for the G6PD deficiency is much lower. This is expected because the value of $c$ is smaller than 1. The disease-specific spatial effects in Figures \ref{sefmjq} and \ref{sefgjq} are very low compared to the shared spatial effect. This can also be seen from the posterior precision estimates for $b_{1}$ and $b_{2}$ compared to that of the shared component in Table \ref{apjq1}. Because the disease-specific spatial effects are smaller than the shared spatial effect, the total spatial effect (that is, the sum of the shared and the specific-spatial effects) is very similar to those for the shared effects, see Figures \ref{tsm10q} and \ref{tsd10q}. The iid random effect for G6PD deficiency, see Figures \ref{iidqjg} and \ref{fig:fig3pmsqj}, is higher than the iid for malaria. This is understandable because the value of Phi for malaria is bigger than the one for G6PD deficiency, which means G6PD deficiency accounts for more iid effect.\\ \\ The relative risks in Figures \ref{rrqmj} and \ref{rrqgj} show similar relative risks as obtained by the separate models. However, observe that the relative risk for G6PD deficiency from the joint model is higher than the one from the separate. This difference between the relative risks is due to borrowing strength from the spatial pattern of malaria through the shared component. The joint quantile model predicts the cases well for both diseases, as can be seen in Figures \ref{pmsqj} and \ref{pgsqj}. The model comparison shows a preference for the joint quantile model over the separate models because the values of DIC and WAIC for the joint quantile are less than the sum of the tests for the separate models. 

\begin{table}[hbt!]
\centering
\begin{tabular}{ccccc}
\hline
\quad & \text{mean}  & \text{sd}  & \text{0.025quant} & \text{0.975quant}  \\
\hline
$m_1$ & 7.852  & 0.634  & 6.58 & 9.12  \\
$m_2$ & 4.245  & 0.285 & 3.669  & 4.81 \\
\hline
\end{tabular}
\caption{The estimated intercepts and the 95 \% credible intervals.}
\label{apjq0}
\end{table}

\begin{figure}[hbt!]
\begin{center}
\includegraphics[scale=0.55]{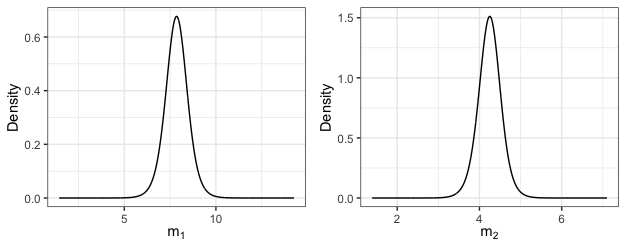}
\caption{The posterior distributions for the intercepts}
\label{apjq00}
\end{center}
\end{figure}

The estimation results of the hyperparameters are given in Tables \ref{apjq1} and \ref{opp}.

\begin{table}[htb!]
\centering
\begin{tabular}{ccccc}
\hline
\quad & \text{mean}  & \text{0.025quant} & \text{0.975quant}  & \text{mode} \\
\hline
$\tau$ & 0.107 & 0.044  & 0.224 & 0.082  \\
$d$ & 1.665  & 0.379 & 4.434  & 0.953 \\
$\tau_{b_{1}}$ & 39.1  & 0.831 & 244.1  & 1.862 \\
$\phi_{b_{1}}$ & 0.321  & 0.021 & 0.827  & 0.056 \\
$\tau_{b_{2}}$ & 1.226  & 0.625 & 2.091 & 1.096 \\
$\phi_{b_{2}}$ & 0.186  & 0.008 & 0.658  & 0.019 \\
$c$ & 0.291  & 0.066 & 0.521 & 0.285 \\
\hline
\end{tabular}
\caption{The estimation of the hyperparameters for low quantile of malaria and high quantile of G6PD deficiency.}
\label{apjq1}
\end{table}

\begin{table}[htb!]
\centering
\begin{tabular}{ccccc}
\hline
\quad & \text{mean}  & \text{0.025quant} & \text{0.975quant}  & \text{mode} \\
\hline
$\tau$ & 3.648  & 1.437  & 7.976 & 2.688  \\
$d$ & 1.819  & 0.423 & 5.258  & 0.977 \\
$\tau_{b_{1}}$ &  11.31 & 5.485 & 24.16  & 8.088 \\
$\phi_{b_{1}}$ & 0.056  & 0 & 0.311  & 0 \\
$\tau_{b_{2}}$ & 131.5  & 55.62 & 326.2  & 82.55 \\
$\phi_{b_{2}}$ & 0.14  & 0 & 0.853  & 0 \\
$c$ & 0.002 & -0.008 & 0.015  & -0.002 \\
\hline
\end{tabular}
\caption{The estimation of the hyperparameters for high quantile of malaria and low quantile of G6PD deficiency.}
\label{opp}
\end{table}

The values for the model choice criteria,  DIC and WAIC, are given in Table \ref{apj2}.

\begin{table}[htb!]
\centering
\begin{tabular}{lcc}
\hline
\quad & \text{DIC}  & \text{WAIC} \\
\hline
\textbf{G6PD Deficiency} & 168 & 164.8 \\
\textbf{Malaria} & 246.8 & 241.4  \\
\textbf{Sum} & 414.8  & 406.2 \\
\textbf{Joint quantile} & 413.6 & 402.2 \\
\hline
\end{tabular}
\caption{Model choice criteria.}
\label{apj2}
\end{table}

\newpage

\begin{figure}
  \includegraphics[width = 5.2cm]{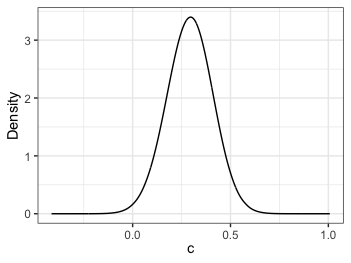}  
\includegraphics[width = 5.2cm]{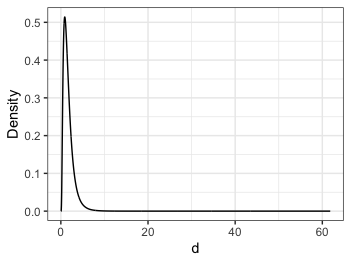}  
  \includegraphics[width = 5.2cm]{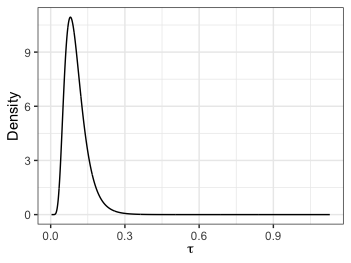} \\ 
  \includegraphics[width = 5.2cm]{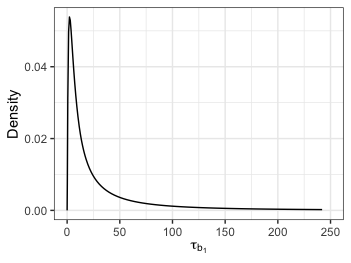}  
\includegraphics[width = 5.2cm]{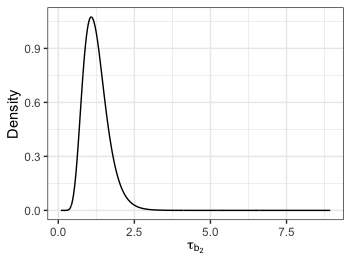}  
\caption{The posterior distributions for the hyperparameters}
\label{Preseions}
\end{figure}

\begin{figure}[hbt!]
\begin{subfigure}{.5\textwidth}
  \centering
  \includegraphics[width=5cm]{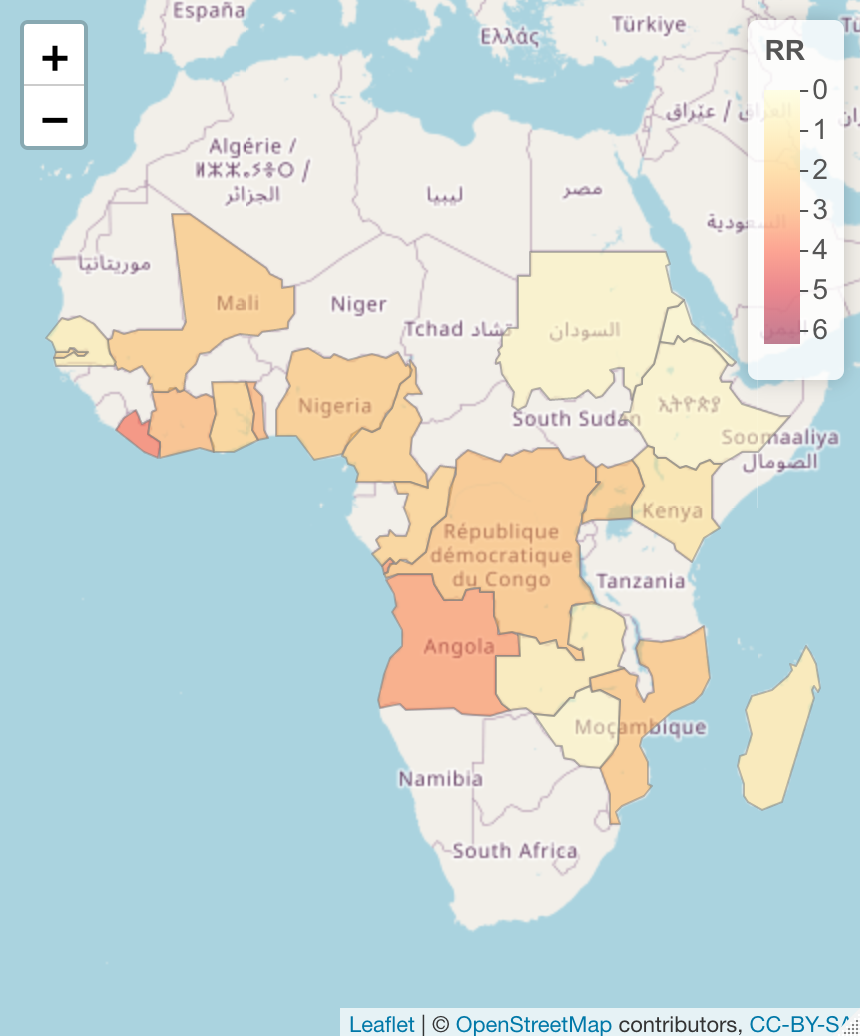}  
  \caption{The relative risk estimate for malaria}
  \label{rrqmj}
\end{subfigure}
\begin{subfigure}{.5\textwidth}
  \centering
\includegraphics[width=5cm]{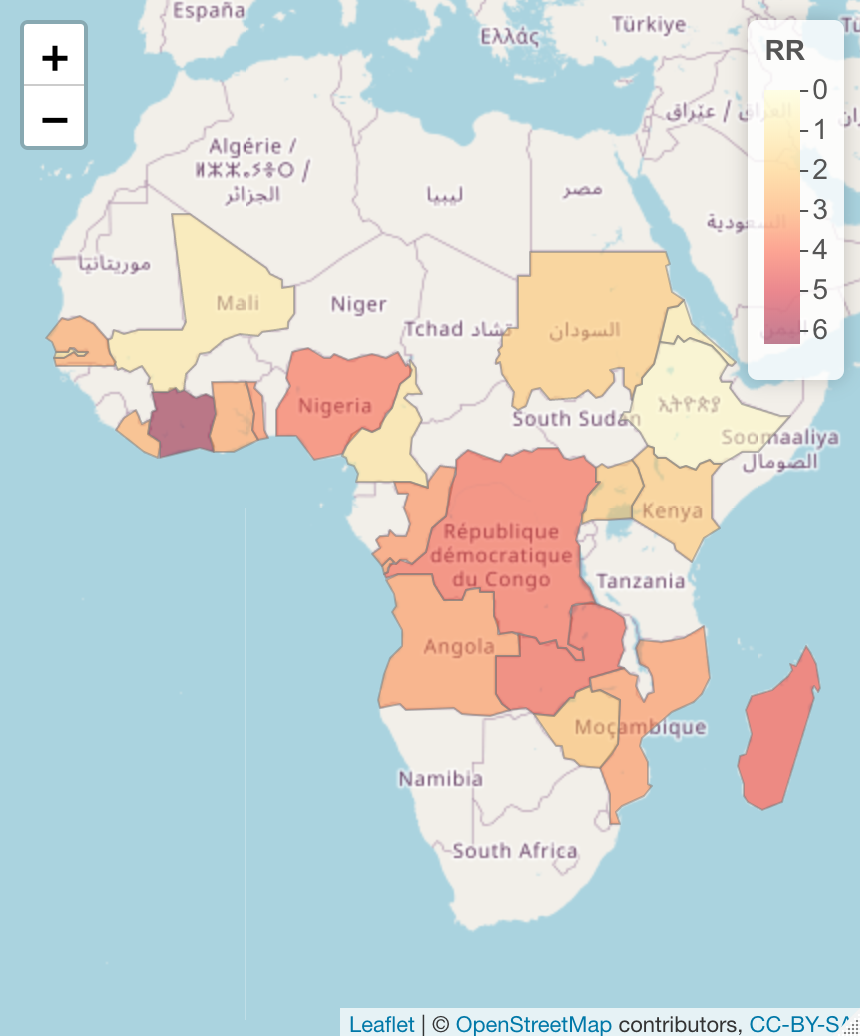}  
  \caption{The relative risk estimate for G6PD}
  \label{rrqgj}
\end{subfigure}
\begin{subfigure}{.5\textwidth}
  \centering
\includegraphics[width=5cm]{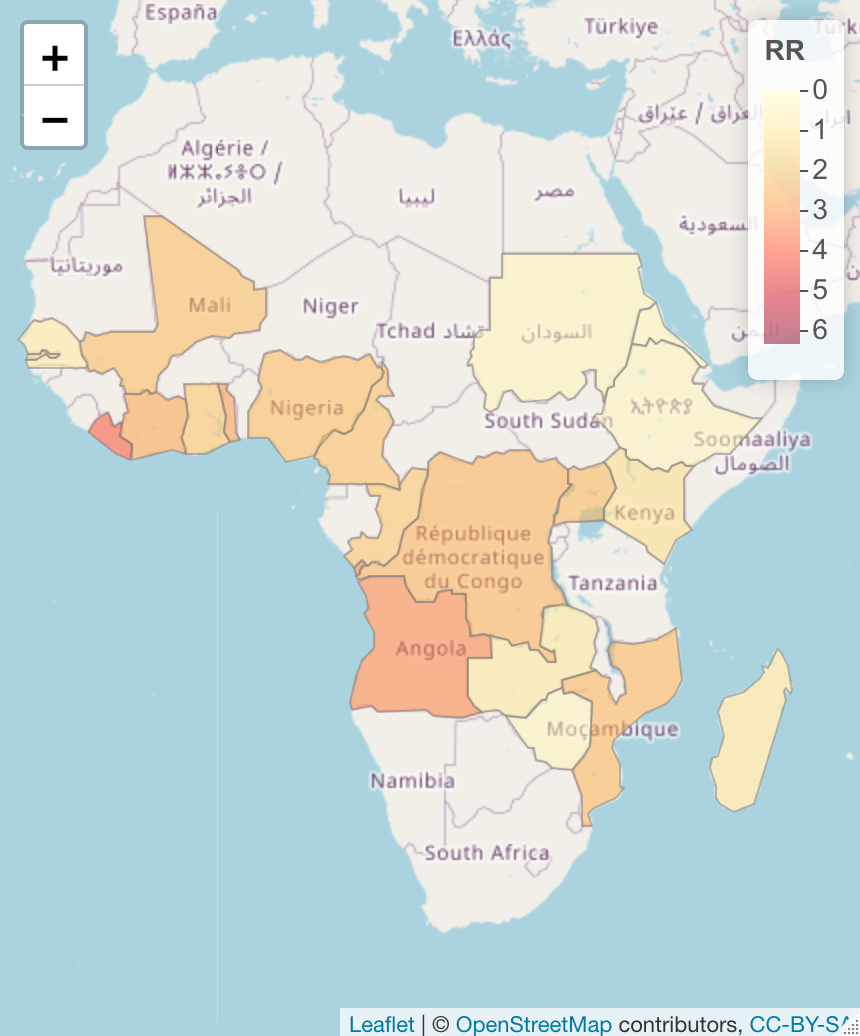}  
  \caption{The relative risk estimate for malaria}
  \label{rrqms}
\end{subfigure}
\begin{subfigure}{.5\textwidth}
  \centering

  \includegraphics[width=5cm]{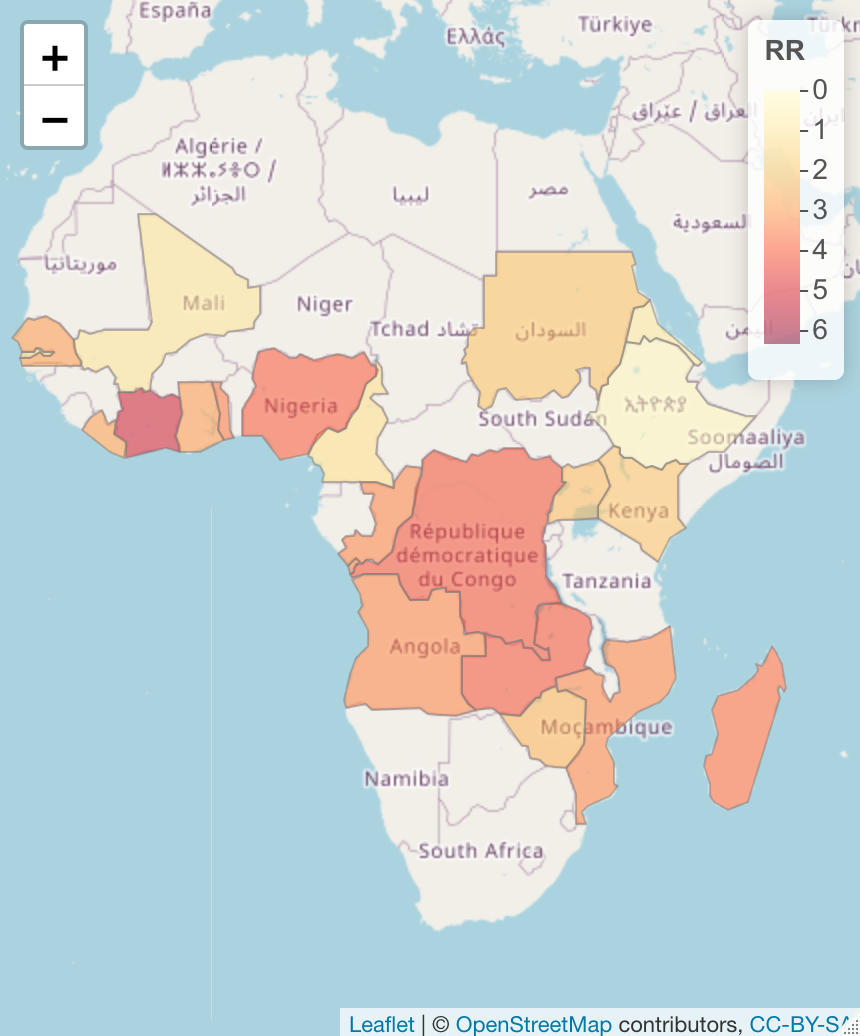}  
  \caption{The relative risk estimate for G6PD}
  \label{rrqgs}
\end{subfigure}
\caption{The estimates from the joint model (top), and the separate models (bottom)}
\label{rrqjs}
\end{figure}

\begin{figure}
\begin{subfigure}{.3\textwidth}
  \includegraphics[width = 5cm]{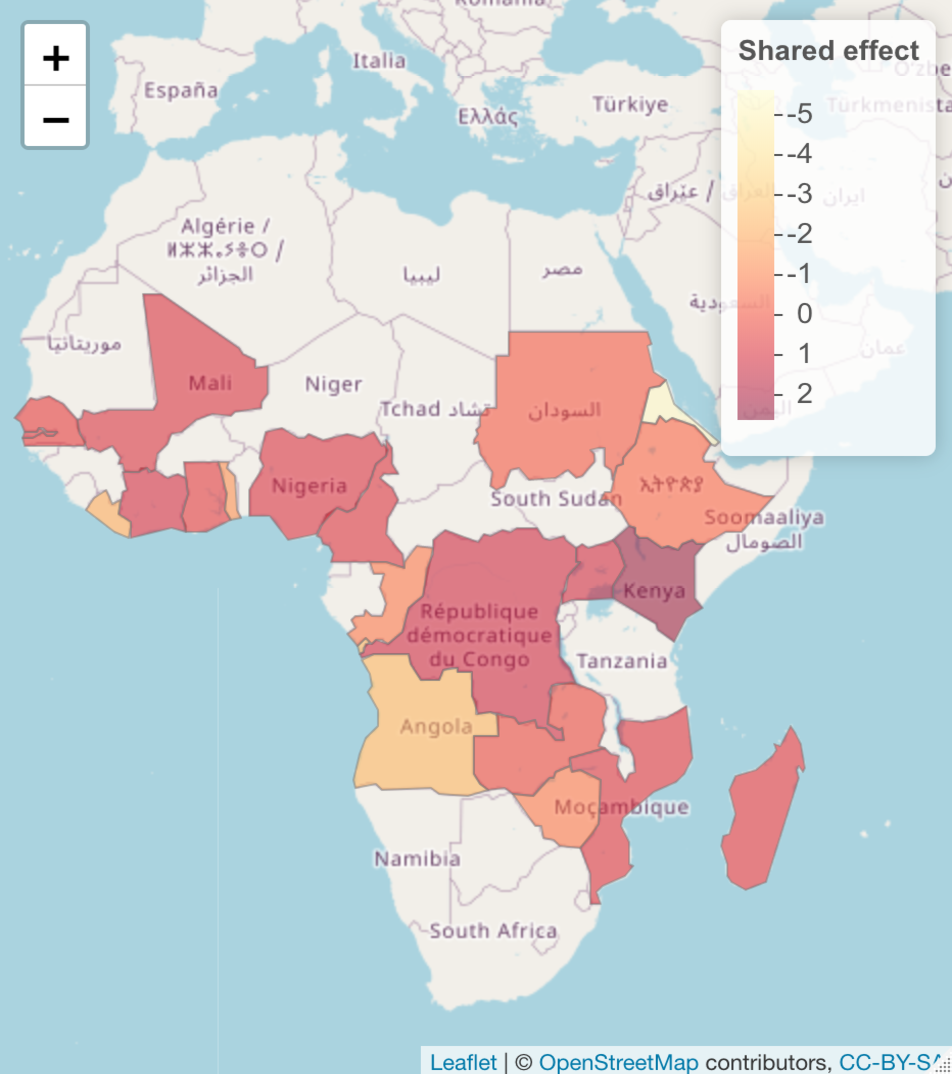}  
  \caption{Shared effect}
  \label{sharedq}
\end{subfigure}
\begin{subfigure}{.3\textwidth}
\includegraphics[width = 5cm]{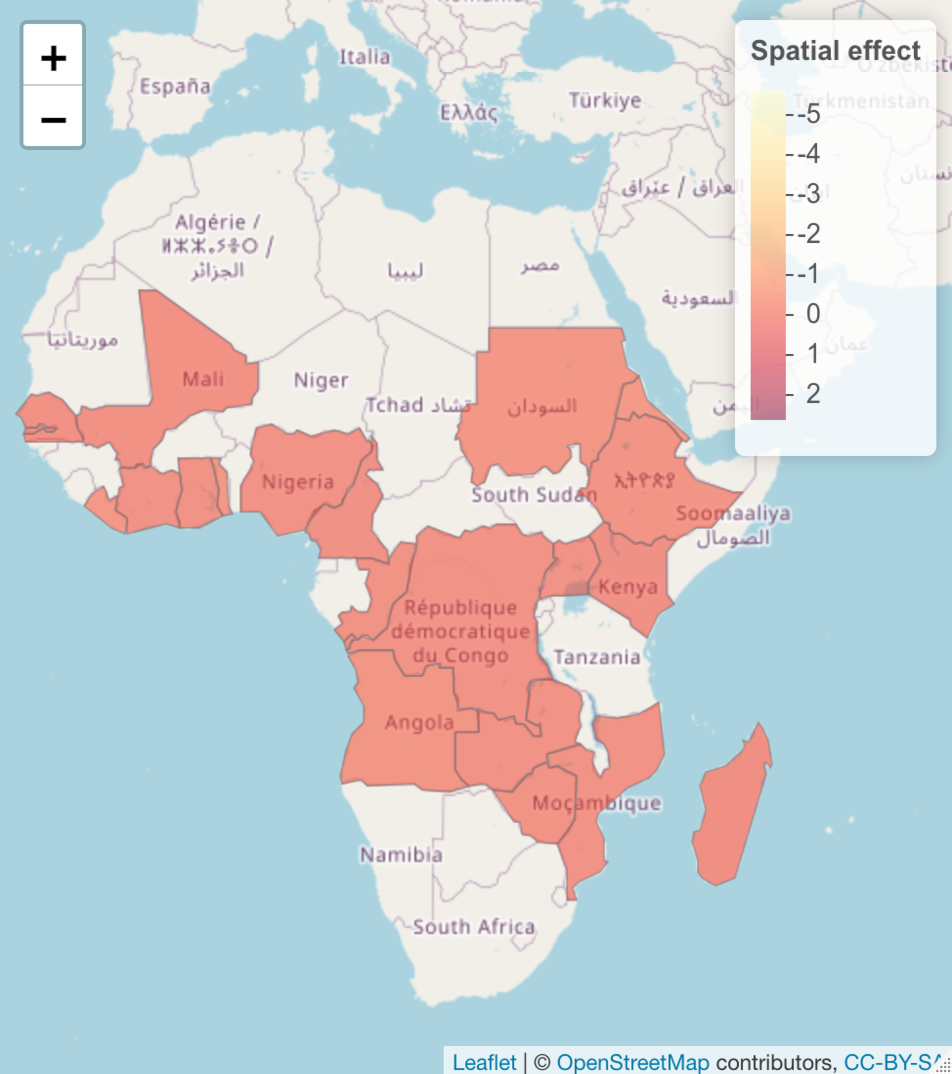}  
  \caption{specific effect}
  \label{sefmjq}
\end{subfigure}
\begin{subfigure}{.3\textwidth}
  \includegraphics[width = 5cm]{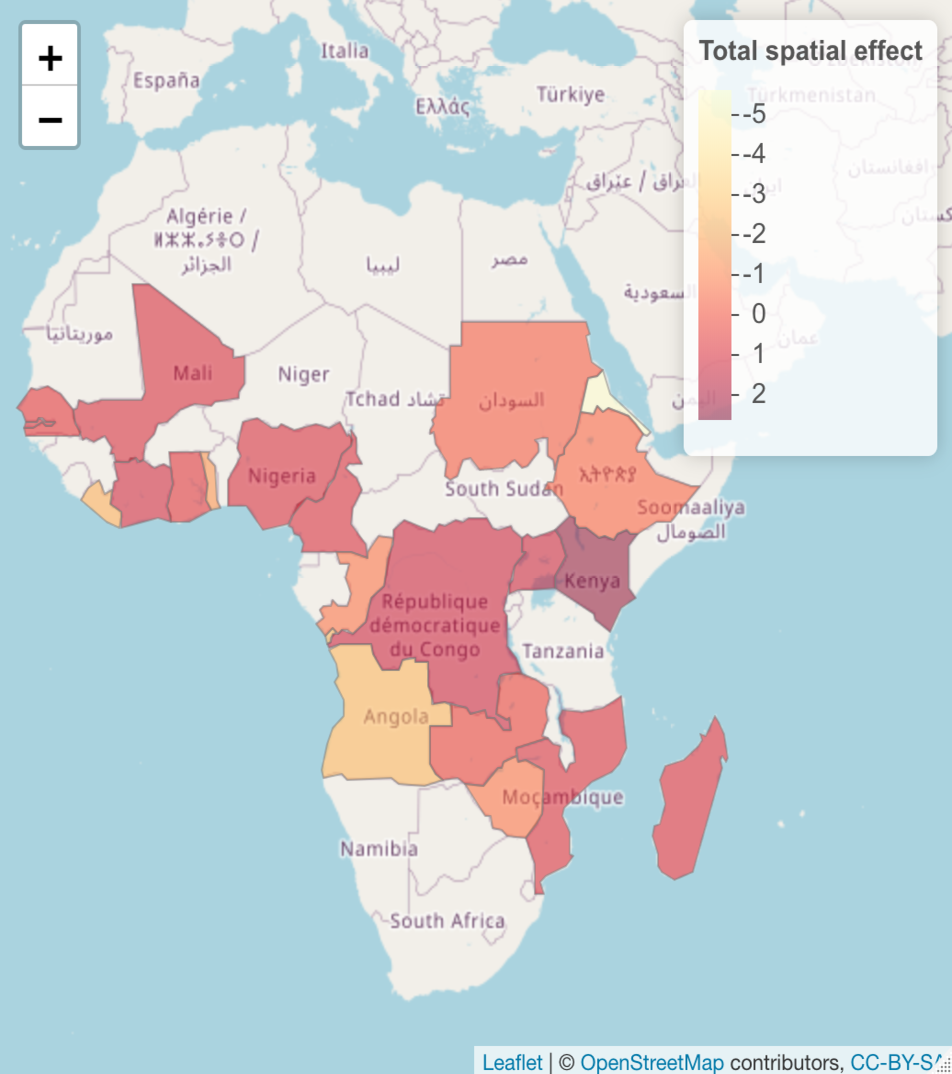}  
  \caption{The total effect}
  \label{tsm10q}
\end{subfigure}

\begin{subfigure}{.3\textwidth}
  \includegraphics[width = 5cm]{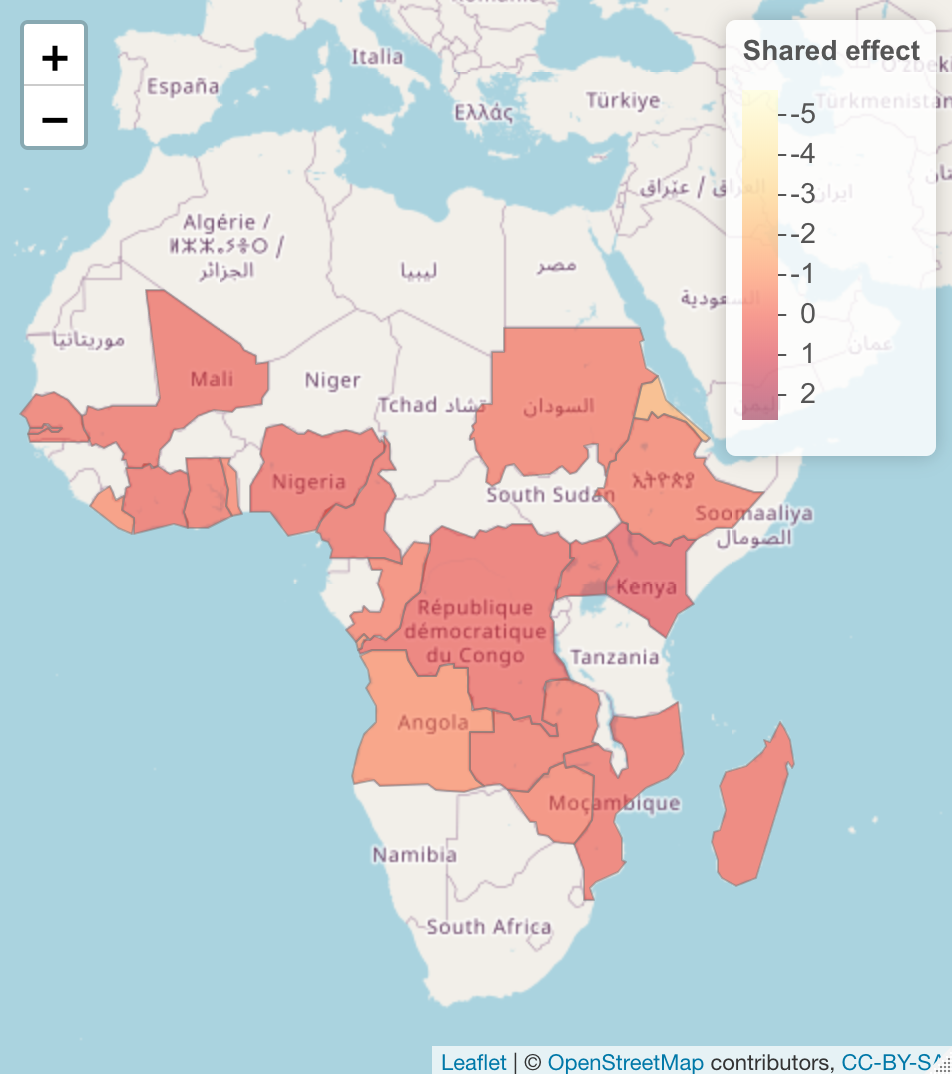}  
\caption{Shared effect}
  \label{sharedq1}
\end{subfigure}
\begin{subfigure}{.3\textwidth}
\includegraphics[width = 5cm]{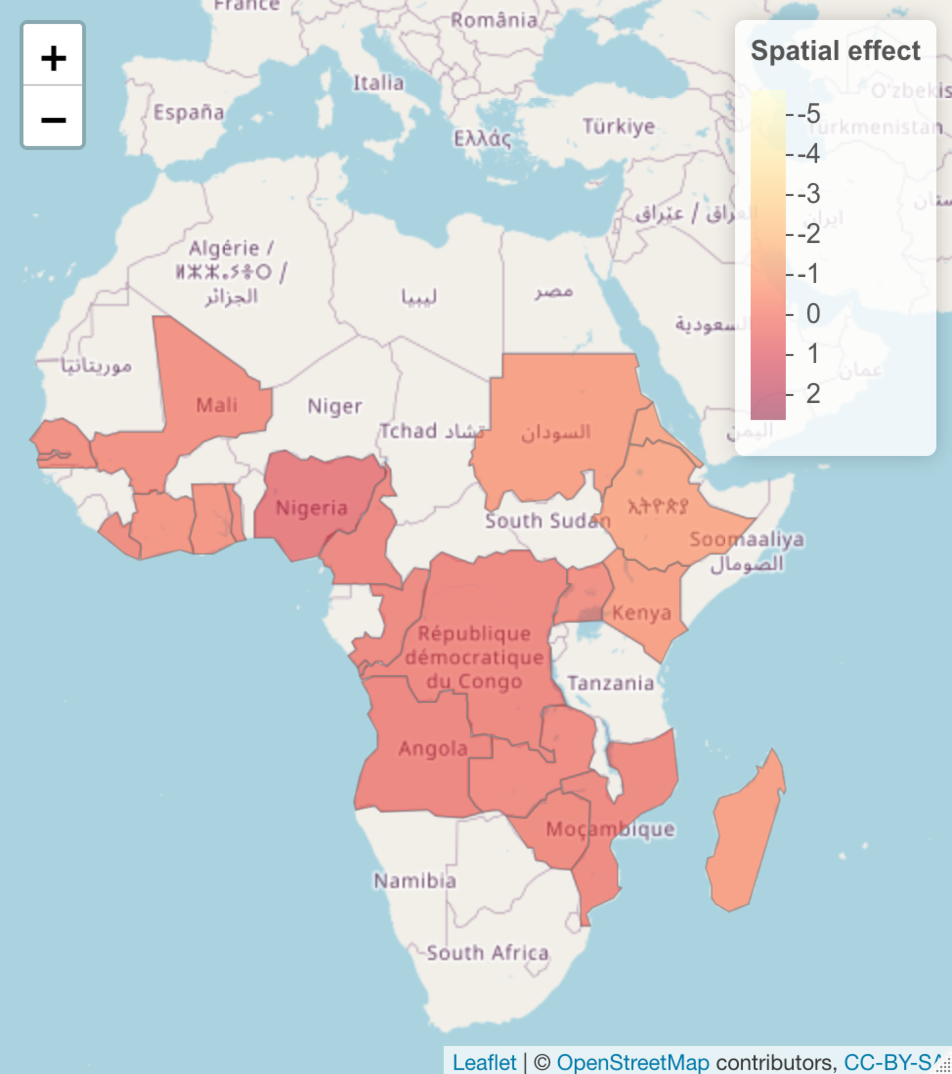}  
  \caption{specific effect}
  \label{sefgjq}
\end{subfigure}
\begin{subfigure}{.3\textwidth}
  \includegraphics[width = 5cm]{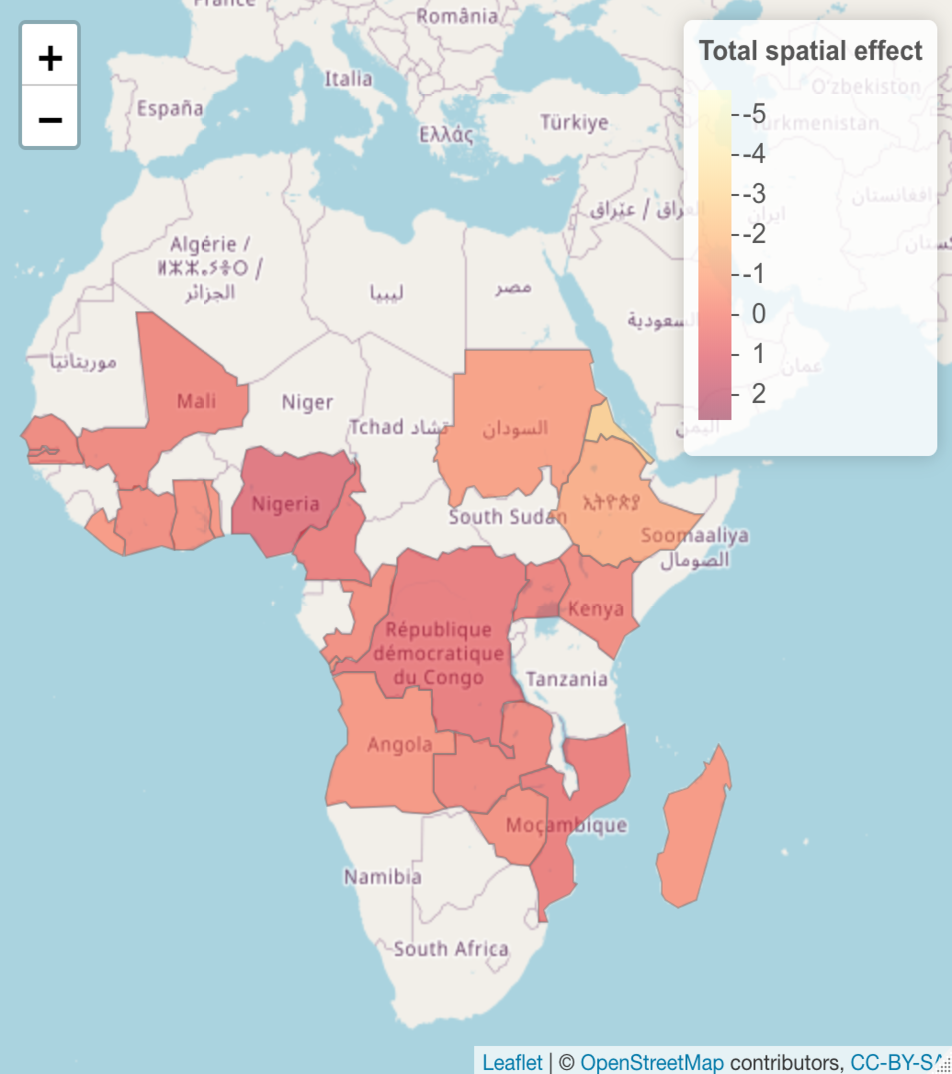}  
  \caption{The total effect}
  \label{tsd10q}
\end{subfigure}
\caption{Posterior estimates of the spatial effects from the joint model for malaria (top) and for G6PD deficiency (bottom)}
\label{jointspatialqant}
\end{figure}

\begin{figure}[hbt!]
\begin{subfigure}{.5\textwidth}
  \centering

  \includegraphics[width=5cm]{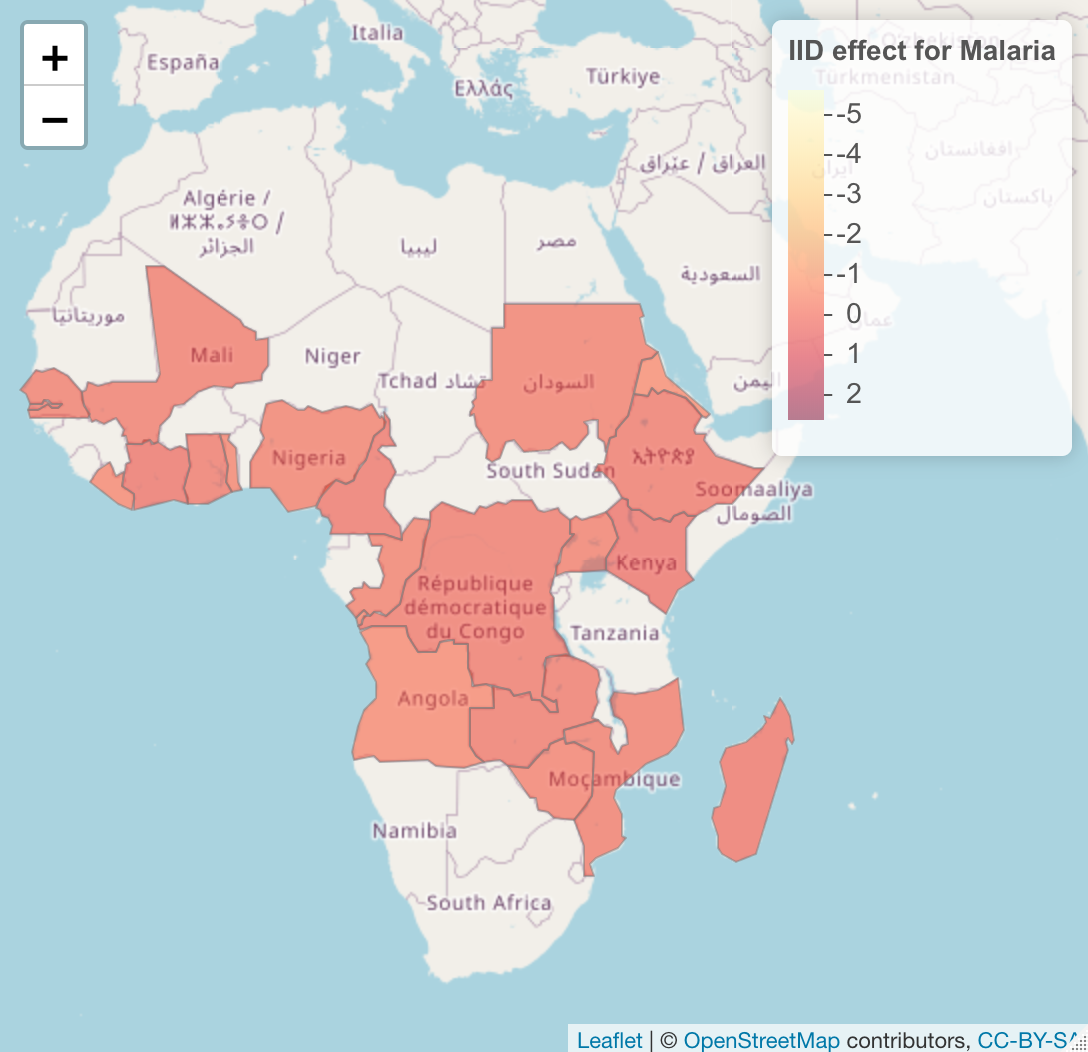}  
  \caption{The iid effect for Malaria}
  \label{iidqjm}
\end{subfigure}
\begin{subfigure}{.5\textwidth}
  \centering

  \includegraphics[width=5cm]{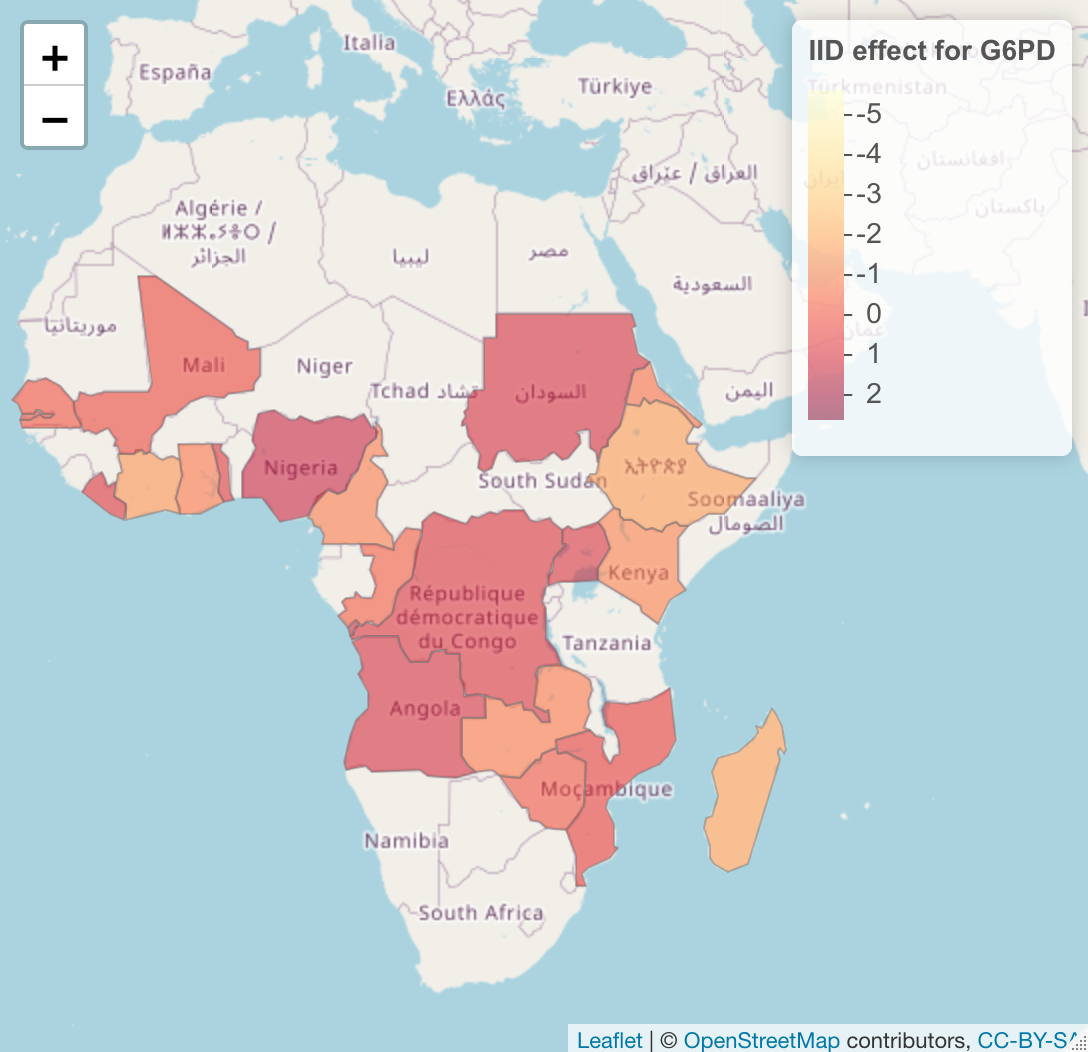}  
  \caption{The iid effect for G6PD}
  \label{iidqjg}
\end{subfigure}
\newline
\begin{subfigure}{.5\textwidth}
  \centering

  \includegraphics[width=5cm]{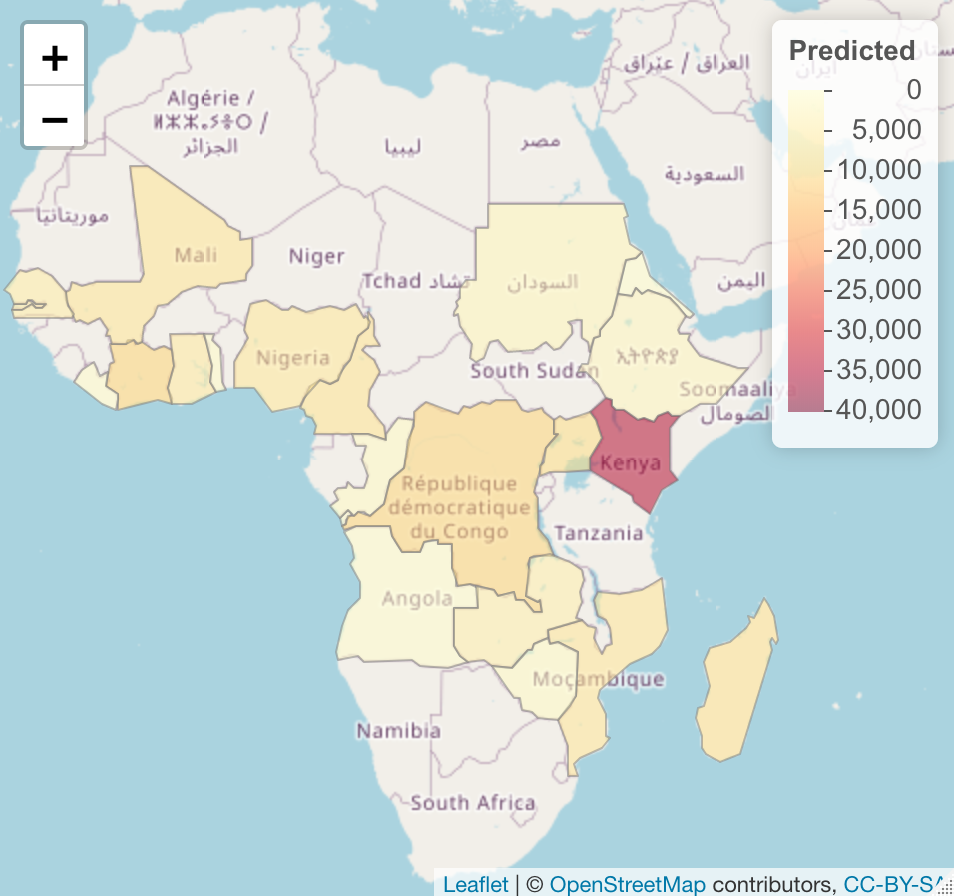}  
  \caption{The predicted cases for Malaria}
  \label{pmsqj}
\end{subfigure}
\begin{subfigure}{.5\textwidth}
  \centering
  
\includegraphics[width=5cm]{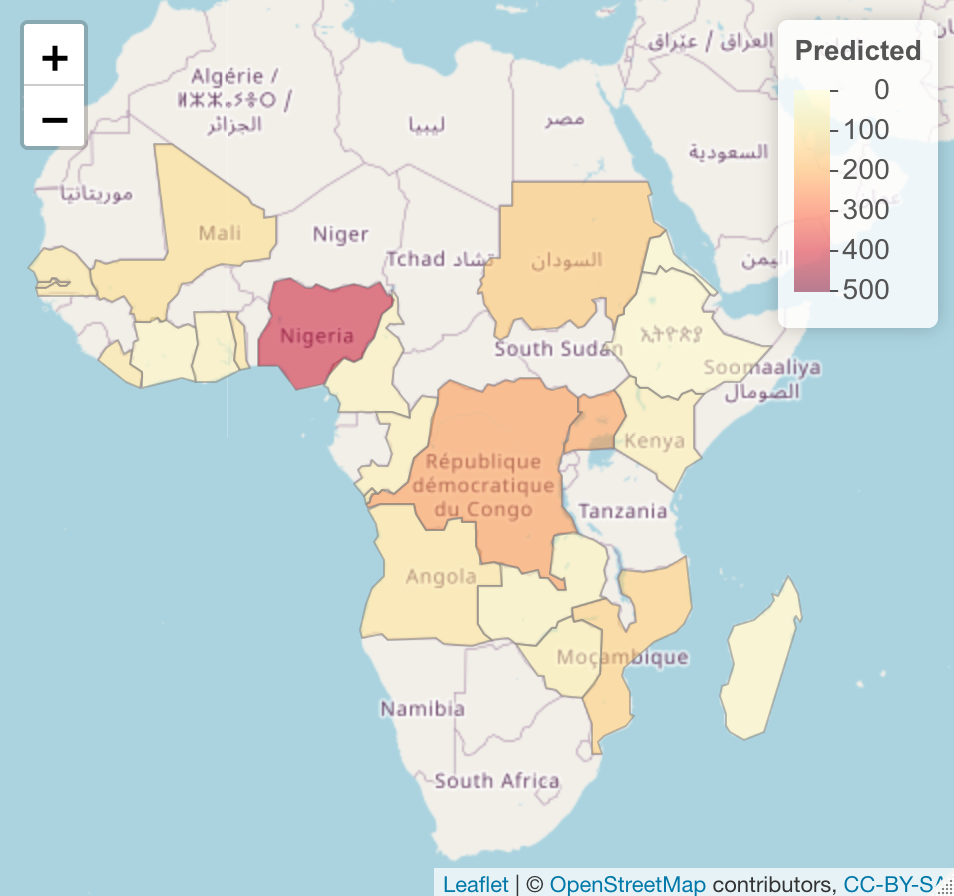} 
  \caption{The predicted cases for G6PD}
  \label{pgsqj}
\end{subfigure}
\caption{The iid effect and predicted cases}
\label{fig:fig3pmsqj}
\end{figure}

\section{Concluding Remarks}
The motivation stemmed from estimating the relative risk of Malaria and G6PD deficiency, jointly, on the African continent. The G6PD deficiency is considered as a resistance against malaria based on anecdotal medical studies (see \cite{beutler1994G6PD} and  \cite{allison1961malaria}). In this case, joint mean disease mapping will not provide the information needed to investigate these initial findings. Therefore, we considered a  joint quantile disease mapping of different quantiles for the diseases. The approach is successful since considering the joint quantile model allows a possible investigation of the correlation between any level of the conditional distributions of the random variables that represents the number of cases, not only the correlation between their means. An advantage of the proposed approach is that the computationally efficient INLA method is used for statistical inference, such as estimating the relative risk. \\ \\
Our main contribution is two-fold. Firstly, we propose a very general joint quantile disease mapping model where the correlation between different quantiles can be inferred and multiple diseases can be considered, together with an efficient computational framework for the inference thereof. Secondly, the significant correlation between a high quantile of G6PD cases and a low quantile of Malaria cases encourages further investigation based on expanded data collection efforts as already underway at the Malaria Atlas Project. This analysis provides a solid statistical framework to the anecdotal findings as remarked by medical professionals, and could underpin future studies in this direction.

\newpage
\bibliographystyle{wileyNJD-AMA}
\bibliography{biblio}

\end{document}